\documentclass[3p, 11pt]{elsarticle}
\usepackage{graphicx,amscd,amsmath,amssymb,verbatim}
\usepackage{amsfonts,epsfig}
\usepackage{mathptmx}
\usepackage{setspace}
\usepackage{epsfig}
\usepackage{url}
\usepackage{multirow}
\usepackage{color}
\usepackage{subfigure}
\usepackage{mathrsfs}
\usepackage{amsfonts}

\begin{document}

\journal{Elsevier}

\begin{frontmatter}

\title{Statistics, distillation, and ordering emergence in a two-dimensional stochastic model of particles in counterflowing streams}

\author{Eduardo Velasco Stock$^{1}$, Roberto da Silva$^{1}$, H. A. Fernandes$%
^{2}$} 

\address{

$^{1}$Instituto de F\'{\i}sica, Universidade Federal do Rio Grande do Sul, 
UFRGS, Porto Alegre - RS, 91501-970, Brasil,\\

$^{2}$Universidade Federal de Goi\'{a}s - UFG, Regional Jata\'{\i}, Jata\'{\i} - GO, 75801-615, Brazil.}

\begin{abstract}

In this paper, we proposed a stochastic model which describes two species of particles moving in counterflow. The model generalizes the theoretical framework describing the transport in random systems since particles can work as mobile obstacles, whereas particles of one species move in opposite direction to the particles of the other species, or they can work as fixed obstacles remaining in their places during the time evolution. We conducted a detailed study about the statistics concerning the crossing time of particles, as well as the effects of the lateral transitions on the time required to the system reaches a state of complete geographic separation of species. The spatial effects of jamming were also studied by looking into the deformation of the concentration of particles in the two-dimensional corridor. Finally, we observed in our study the formation of patterns of lanes which reach the steady state regardless the initial conditions used for the evolution. A similar result is also observed in real experiments involving charged colloids motion and simulations of pedestrian dynamics based on Langevin equations, when periodic boundary conditions are considered (particles counterflow in a ring symmetry). The results obtained through Monte Carlo numerical simulations and numerical integrations are in good agreement with each other. However, differently from previous studies, the dynamics considered in this work is not Newton-based, and therefore, even artificial
situations of self-propelled objects should be studied in this first-principle modeling.

\end{abstract}

\end{frontmatter}

\tableofcontents

\setlength{\baselineskip}{0.7cm}

\section{Introduction}

In condensed matter physics, the study of the transport of particles in
random environments under quenched or annealed scenarios has a huge number
of applications such as the capture/decapture of electrons in the
micro/nano/meso devices \cite{Machlup1954,Kirton1989,noisesemiconductors},
the random motion of molecular motors in the cellular transport (see, for
example, Ref. \cite{Goldman}), the erratic motion of molecules in
cromatographic columns \cite{Cromatograph} and many others. Among these
problems, one in particular called our attention: \textit{the counterflowing
streams of particles}, which can appear in many contexts due to its
importance, for instance, in the separation of chemical products \cite%
{Li2014}, pedestrian dynamics \cite%
{HelbingNature,Helbing1995,Gawronski2011,Peng2011,Nakayama2007,rdasilva2015,Pinho2016}%
, and band formation in mixtures of oppositely charged colloids \cite%
{Vissers2011,VissersPRL2011}.

Recently, more precisely in the context of pedestrian dynamics, some authors
have studied the problem by considering the equation of motion for each
particle as a Langevin-like equation \cite{Pinho2016} which uses the known
and successful approach of social force developed by Helbing and Moln\~{A}%
\textexclamdown r in 1995 \cite{Helbing1995}. Basically, such approach
defines that the dynamics depends on deterministic and stochastic forces.
They considered a stokesian drag force that imposes velocities on the
particles, an interaction force (repulsion) among particles, an interaction
force of particles with the system boundaries, and a Gaussian noise term
(based on the existence of an arbitrary motion in a crowd). The authors \cite%
{Pinho2016} have shown that asymmetrically shaped walls in a corridor with
pedestrian counterflow can surprisingly organize the flow of pedestrians in
two only opposite directions (\textquotedblleft keep-left\textquotedblright\
behavior). In pedestrian dynamics, it is important to observe that lane
formation phenomenon in pedestrian counterflow were observed via numerical
results in other situations. For instance, in Ref. [\cite{Xiong2011}], the
simulations based on optimal path-choice strategy showed that segregation is
associated with the minimization of the travel time to reach the ordering
state. Similarly, an extended lattice-gas model under periodic boundary
conditions is also able to lead to such lane patterns \cite{Xiang2012}. Such
results are experimentally corroborated by real situations as shown by Kretz
et al. \cite{Kretz2006} by using pedestrian counterflow experiment in a
corridor of width 2 m and 67 participants.

In addition, based on the concept of clannish random walks, Montroll and
West \cite{MontrollandWest} have given an alternative approach for the study
of these problems of counterflowing streams of particles in a special kind 
of mean-field regime. By making the necessary changes and
adaptations, we proposed in Ref. [\cite{rdasilva2015}] a one-dimensional
stochastic dynamics based on an embryonic idea of those authors by
substituting the effects of social forces and other contributions by the
following assumptions: (a) from a microscopic point of view, more than one
particle can occupy the same cell and (b) each particle of one species jumps
to a neighbor cell with probability which depends on the local concentration
of particles of the other species.

By following the experimental results of Vissers et al. \cite%
{Vissers2011,VissersPRL2011} about patterns produced by oppositely charged
particles under action of electric fields, we are wondering about the
minimal ingredients necessary to produce lane patterns in a stochastic
dynamics based on occupation of cells. So, in this paper we propose a
two-dimensional model in order to describe the motion of two species of
particles ($A$ and $B$) moving in opposite directions. In our model, the
evacuation probabilities of a given species located on a cell change
according to the concentration of the other species on this same cell at
that time. This is a general and simple way to model the possible
interactions between these particles. Additionally, our model was motivated
by the pedestrian dynamics and the stationary cases are compared with
alternative models as, for example, that one presented in Ref. [\cite%
{Pinho2016}].

The generality of our stochastic modeling offers many alternatives of
scenarios as suggested in Fig. \ref{Figure:Scheme_of_dynamics}, and
depending on the choice of the parameters, we can apply the model for
different situations:

\begin{figure}[t]
\begin{center}
\includegraphics[width=0.8\columnwidth]{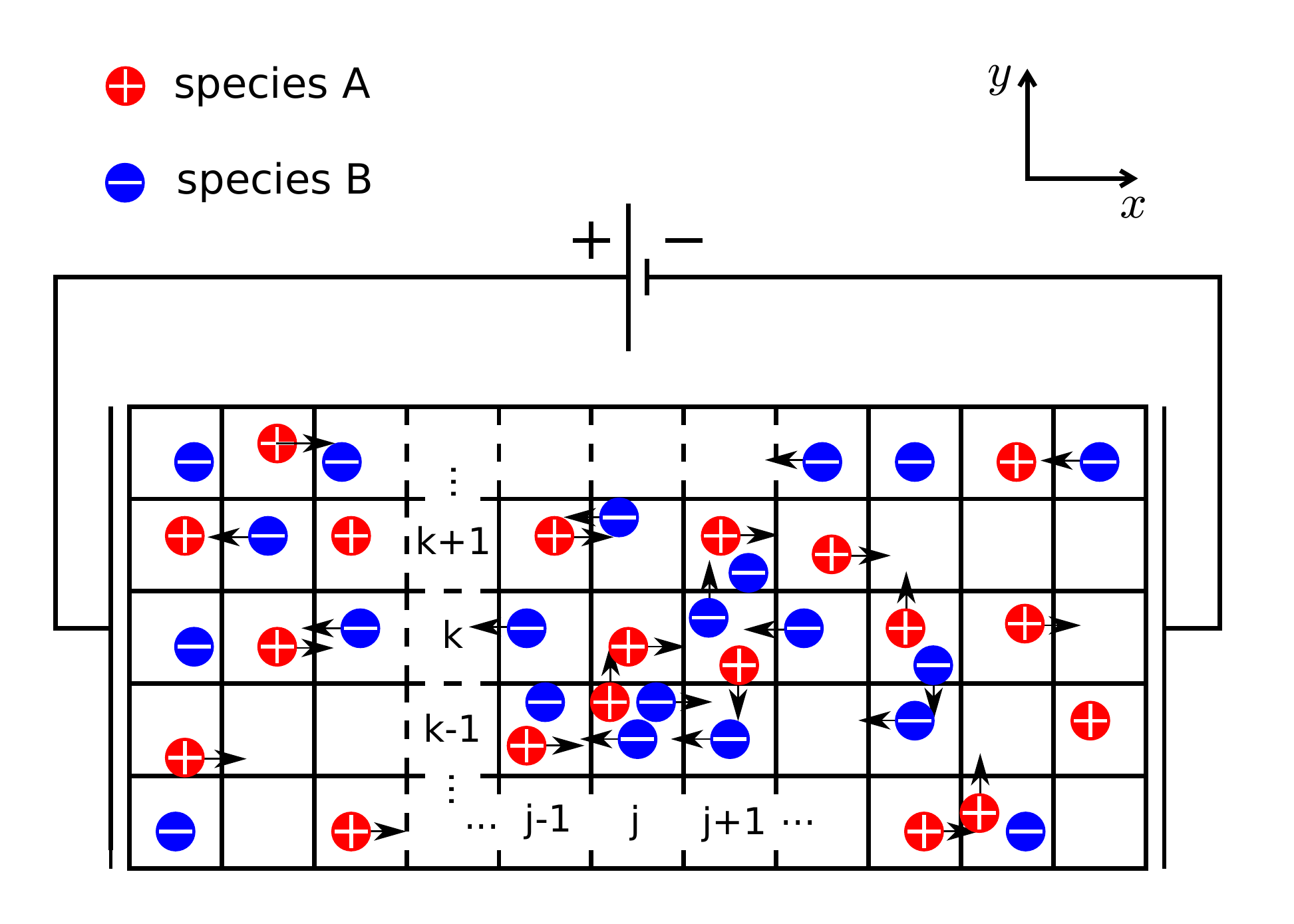}
\end{center}
\caption{Scheme of the stochastic model of two species of particles in a
two-dimensional lattice where the particles can transit to the neighbor
cells. In this modeling, the particles $B$ are able to move or can remain in
fixed cells considering the flux of the target particles $A$. In a first
case (annealed scenario) we are considering conterflowing streams of
particles of species $A$ against the species $B$. On the other hand, whe can
consider the particles of species $B$ working as fixed obstacles (quenched
scenario). In this last case, the motion of particles $A$ is an interesting
example of ordered transport of particles with fixed impurities spatially
distributed in the environment. The analogy with a charged system and an
electric field applied along the longitudinal direction is explored in this
figure.}
\label{Figure:Scheme_of_dynamics}
\end{figure}

\begin{enumerate}
\item \textbf{Annealed scenario}: The species move against each other, for
example, $A$ to the right and $B$ to the left composing two counterflowing
streams of particles. The two streams interact affecting each other. In a
cell, the concentration of particles of a given species changes as time
evolves and therefore the transition probabilities of the another species to
that cell also change according to this concentration. We can imagine
oppositely charged particles under action of electric fields in the context
of granular materials \cite{Vissers2011,VissersPRL2011} etc. So, we can
observe jamming effects that emerge due to the resistance of the oppositely
charged particles moving in opposite direction, since the motion depends on
their concentration per cell. Another interesting example is the motion of
the two counterflowing streams of pedestrians in a corridor. In that case,
the self-propulsion of the pedestrians into a preferential direction plays
the role of the electric field \cite{rdasilva2015,Pinho2016};

\item \textbf{Quenched scenario}: The particles $B$ remain in fixed
positions chosen at random at the beginning of the time evolution. 
In this case, the particles of the species $B$ work as obstacles for the
particles of species $A$. One can imagine the particles $B$ as fixed
impurities in a material and the model can be thought as a traditional
random (but directed) walk of particles in a general transport phenomena,
exactly as happens in the electrons transport in the interfaces of a
semiconductor, chromatograph process of molecules, motion in dopped
materials and many others (see for example \cite%
{noisesemiconductors,Cromatograph}).
\end{enumerate}

In this paper, we describe the statistical fluctuations and the relaxation
process considering different initial conditions in both scenarios
considering a rectangular lattice that may or may not have periodic boundary
conditions depending on the studied problem. We also show an approach that
considers both Monte Carlo (MC) simulations and Numerical Integration (NI)
of recurrence equations of the flows of particles along the lattice. The
partial differential equations that govern the dynamics at continuum limit
are also deduced.

In the annealed scenario we can separate our contribution in three main
parts:

\begin{enumerate}
\item Statistical characterization of spatial distribution of particles
which considers one strip of each species located initially at the opposite
extremities of the corridor;

\item By considering the two species randomly mixed in the lattice at the
beginning of the evolution, we analyze the fluctuations in the times needed
to the complete separation of the two species (distillation);

\item We consider the emergence of ordering by longitudinal bands
considering periodic boundary conditions (or simply that the two species can
rotate in a ring). We defined an interesting order parameter to quantify the
appearance of bands considering the interaction of the counterflowing
streams of the two different species.
\end{enumerate}

On the other hand, in the quenched scenario, we concentrate our analysis in
measuring the fluctuations on the crossing times of the target particles
considering different densities of the fixed particles (obstacles). We
observe an interesting crossover for these crossing times that are normally
distributed (a limit case of a negative binomial distribution) at low
concentration of obstacles and become distributed according to a power law
in a transient situation corresponding to an intermediate concentration of
obstacles. Finally, exponential tails appear after a high concentration of
particles which makes the system very stagnant.

In the next section we explain the details of the modeling and definitions
of the order parameters which describe the cellular segregation of the
particles in the different studies. In section \ref{SecResults:I} we present
the results related to the situation where one of the species is not able to
moove, i.e., the particles of that species is fixed in the cells and work as
obstacles for the oriented transport of the target particles (the particles
of the other species). In section \ref{SecResults:II}, we present our main
results when the two species are oriented in opposite directions forming
counterflowing streams of particles. We finally present some summaries and
our conclusions in section \ref{Sec:Conclusions}.

\section{Methods and modeling}

\label{Sec:Methods_and_Modeling}

In this model, we consider a corridor with linear dimensions $L_{x}$ and $%
L_{y}$, the width and length dimensions, respectively, so that we have a
total of $L_{x}L_{y}$ cells in this lattice. We take into account particles
of species $A$ ($+$) that move preferably to the left and particles of
species $B$ ($-$) moving preferably to the right according to, for example,
an application of a uniform field (which is supposed to be constant). We
denote $\rho_{A}(j,k)=\frac{c_A(j,k)}{c_A(j,k)+c_B(j,k)}$ and $\rho_B(j,k)=%
\frac{c_B(j,k)}{c_A(j,k)+c_B(j,k)}$ as the concentration of particles $A$
and $B$, respectively, in the cell $(j,k)$ with $j=1,\cdots,L_{x}$ and $%
k=1,\cdots,L_{y}$. Here, $c_A$ and $c_B$ are, respectively, the total number
of particles of the species $A$ and $B$ in the cell.

In the more general context (annealed system), we denote $p$ as the
probability that directs the particles of both species, which depends on the
applied field. So, the interaction occurs when the counterflowing particles
visit the same cells. For the species $A$ this interaction is defined by
(the same analysis present below is applied to the species $B$): 
\begin{equation}
\begin{array}{l}
\Pr^{(A)}(j\rightarrow j+1,k\rightarrow k)=p-\alpha \rho _{B}(j,k) \\ 
\\ 
\Pr^{(A)}(j\rightarrow j+1,k\rightarrow k\pm 1)=\beta _{\perp }\rho _{B}(j,k)
\\ 
\\ 
\Pr^{(A)}(j\rightarrow j-1,k\rightarrow k)=\beta _{\parallel }\rho _{B}(j,k)
\\ 
\\ 
\Pr^{(A)}(j\rightarrow j,k\rightarrow k)=1-p+(\alpha -\beta )\rho _{B}(j,k)%
\end{array}
\label{Eq.Main_equation}
\end{equation}%
where $\beta =\beta _{\perp }+\beta _{\parallel }$. Here $\alpha $ is
related to the frontal resistance that particles offer to the reference
particles as considered in Ref. [\cite{rdasilva2015}]. The particles can
move to the right or to the left with the same probability $\beta _{\perp
}\rho _{A}(j,k)$. We introduce an elastic effect that promotes the return of
the particle to the previous cell with probability $\beta _{\parallel }\rho
_{A}(j,k)$.

By denoting $n_{j,k,l}$ the number of target particles in the cell ($j,k$)
at the instant $l$, the first approach leads to the recurrence equation 
\begin{equation}
\begin{array}{l}
n_{j,k,l+1}=n_{j,k,l}-p(n_{j,k,l}-n_{j-1,k,l}) \\ 
\\ 
+\alpha \left[ \left( 1-\frac{n_{j,k,l}}{N_{j,k,l}}\right) n_{j,k,l}-\left(
1-\frac{n_{j-1,k,l}}{N_{j-1,k,l}}\right) n_{j-1,k,l}\right] \\ 
\\ 
+\beta _{\parallel }\left[ \left( 1-\frac{n_{j+1,k,l}}{N_{j+1,k,l}}\right)
n_{j+1,k,l}-\left( 1-\frac{n_{j,k,l}}{N_{j,k,l}}\right) n_{j,k,l}\right] \\ 
\\ 
+\beta _{\perp }\left[ \left( 1-\frac{n_{j,k+1,l}}{N_{j,k+1,l}}\right)
n_{j,k+1,l}+\left( 1-\frac{n_{j,k-1,l}}{N_{j,k-1,l}}\right)
n_{j,k-1,l}\right. \\ 
\\ 
-\left. 2\left( 1-\frac{n_{j,k,l}}{N_{j,k,l}}\right) n_{j,k,l}\right] .%
\end{array}
\label{Eq:recurrence_equations}
\end{equation}%
where $N_{j,k,l}=n_{j,k,l}+m_{j,k,l}$ is the total of particles in the cell (%
$j,k$) at the instant $t=l\Delta $ and $m_{j,k,l}$ is the number of opposite
particles in the same cell and instant.

This corresponds to the discretizations of the EDP: 
\begin{equation}
\frac{\partial c_{A}}{\partial t}=-k_{1}\frac{\partial c_{A}}{\partial x}%
+k_{2}\frac{\partial }{\partial x}\left( \frac{c_{A}c_{B}}{c_{A}+c_{B}}%
\right) +k_{3}\frac{\partial ^{2}}{\partial y^{2}}\left( \frac{c_{A}c_{B}}{%
c_{A}+c_{B}}\right) ,  \label{Eq:EDP}
\end{equation}%
where $k_{2}=\lim_{a,b,\tau \rightarrow 0}\frac{a}{\Delta }(\alpha +\beta
_{\parallel })$ and $k_{3}=\lim_{a,b,\tau \rightarrow 0}\frac{b^{2}}{\Delta }%
\beta _{\perp }$. The parameters $a$ and $b$ are respectively the dimensions
of the cell, and $\tau $ is the time interval between the transitions.
Although numerical properties of Eq. (\ref{Eq:EDP}) deserves attention, in
order to compare with MC simulations, we can simply make $a=b=\Delta =1$
since each time unit corresponds to one MC step and our intention is simply
to integrate the recurrence equations.

The quenched scenario is obtained simply by making $\Pr^{(B)}(\cdot
\rightarrow \cdot )$ identically equal to zero in Eq. (\ref{Eq.Main_equation}%
) and $\Pr^{(A)}(\cdot \rightarrow \cdot )$ calculated exactly as in the
annealed scenario, i.e., by following Ref. [\ref{Eq.Main_equation}].

\subsection{The order parameters: cellular, transversal, and longitudinal
segregation}

In this work, it is interesting to use order parameters that quantitatively
measure the patterns of separation of the particles in the lattice. For
example, in models of charged colloids \cite{Vissers2011} as well as
pedestrian models \cite{Pinho2016} we can observe interesting band formation
patterns. Such bands must be formed taking into account the direction of the
applied field that interact with the particles in the lattice, as well as
the boundary conditions of the problem. In a general way, we can define the
cellular segregation in a two-dimensional system independently of the
direction at the time $t=l\tau $: 
\begin{equation}
\begin{array}{lll}
\Phi _{cell}^{l} & = & \frac{\sum_{j=1,k=1}^{Lx,Ly}|n_{j,k,l}-m_{j,k,l}|}{%
\sum_{j=1,k=1}^{Lx,Ly}(n_{j,k,l}+m_{j,k,l})} \\ 
&  &  \\ 
& = & \frac{1}{N}\sum_{j=1,k=1}^{Lx,Ly}|n_{j,k,l}-m_{j,k,l}|\text{,}%
\end{array}
\label{eq_cellular_order_parameter}
\end{equation}%
where $N\equiv N_{A}+N_{B}$ is the total number of particles in the lattice.
This order parameter measures how the species are segregated by each other
per cell, or simply how much the species are separated per cells. On the
other hand, if we are looking for band formation (transversal or
longitudinal) which are particular cases of segregation, we must define two
particular segregation order parameters:

1) \textbf{The transversal segregation order parameter} 
\begin{equation}
\begin{array}{lll}
\Phi^{l}_{\perp } & = & \frac{\sum_{j=1}^{Lx}|%
\sum_{k=1}^{Ly}(n_{j,k,l}-m_{j,k,l})|}{%
\sum_{j=1,k=1}^{Lx,Ly}(n_{j,k,l}+m_{j,k,l})} \\ 
&  &  \\ 
& = & \frac{1}{N}\sum_{j=1}^{Lx}|n_{j,l}-m_{j,l}|,%
\end{array}
\label{eq_trans_order_parameter}
\end{equation}

2) \textbf{The longitudinal segregation order parameter} 
\begin{equation}
\begin{array}{ccc}
\Phi^{l}_{\parallel } & = & \frac{\sum_{k=1}^{Ly}|%
\sum_{j=1}^{Lx}(n_{j,k,l}-m_{j,k,l})|}{%
\sum_{j=1,k=1}^{Lx,Ly}(n_{j,k,l}+m_{j,k,l})} \\ 
&  &  \\ 
& = & \frac{1}{N}\sum_{k=1}^{Ly}|n_{k,l}-m_{k,l}|\text{.}%
\end{array}
\label{eq_long_band_order_parameter}
\end{equation}

Such proposed parameters are used to describe the dynamics relaxation of our
model in two different contexts: (1) distillation of particles and (2) band
formations when the particles are looping in a ring or simply by considering
that the particles have periodic boundary conditions in the longitudinal
direction.

\section{Results I: species B as fixed obstacles -- Quenched scenario}

\label{SecResults:I}

First of all, we start our study observing the statistical analysis of the
spatial and temporal properties of the particles. So, we consider the
particles of the species $A$ initially disposed as a transversal strip, on
the far left of the two dimensional lattice, which means 
\begin{equation*}
n_{A}(j,k;t=0)=\left\{ 
\begin{array}{lll}
\frac{N_{A}}{L_{y}} & \text{if} & 
\begin{array}{l}
j=1\text{,} \\ 
k=1,\cdots ,L_{y}%
\end{array}
\\ 
&  &  \\ 
0 & \text{if} & 
\begin{array}{l}
2<j<L_{x}\text{,\ } \\ 
k=1,\cdots ,L_{y}%
\end{array}%
\end{array}%
\right. 
\end{equation*}%
while the particles of the species $B$, at rest, are uniformly distributed
in the corridor, i.e., $n_{B}(j,k;t=0)\approx \frac{N_{B}}{L_{x}L_{y}}$ .
This assumption corresponds to the second scenario described in the
introduction (the particles of the species $B$ are fixed obstacles to the
particles $A$). So, we measured the crossing time $\tau $ (the necessary
time for the particles of the species $A$ cross the corridor) in order to
explore the effects of some parameters such as $N_{A}$, $N_{B}$, $\alpha $, $%
\beta _{\parallel }$, and $\beta _{\perp }$. 
\begin{figure}[tbp]
\begin{center}
\includegraphics[width=\columnwidth]{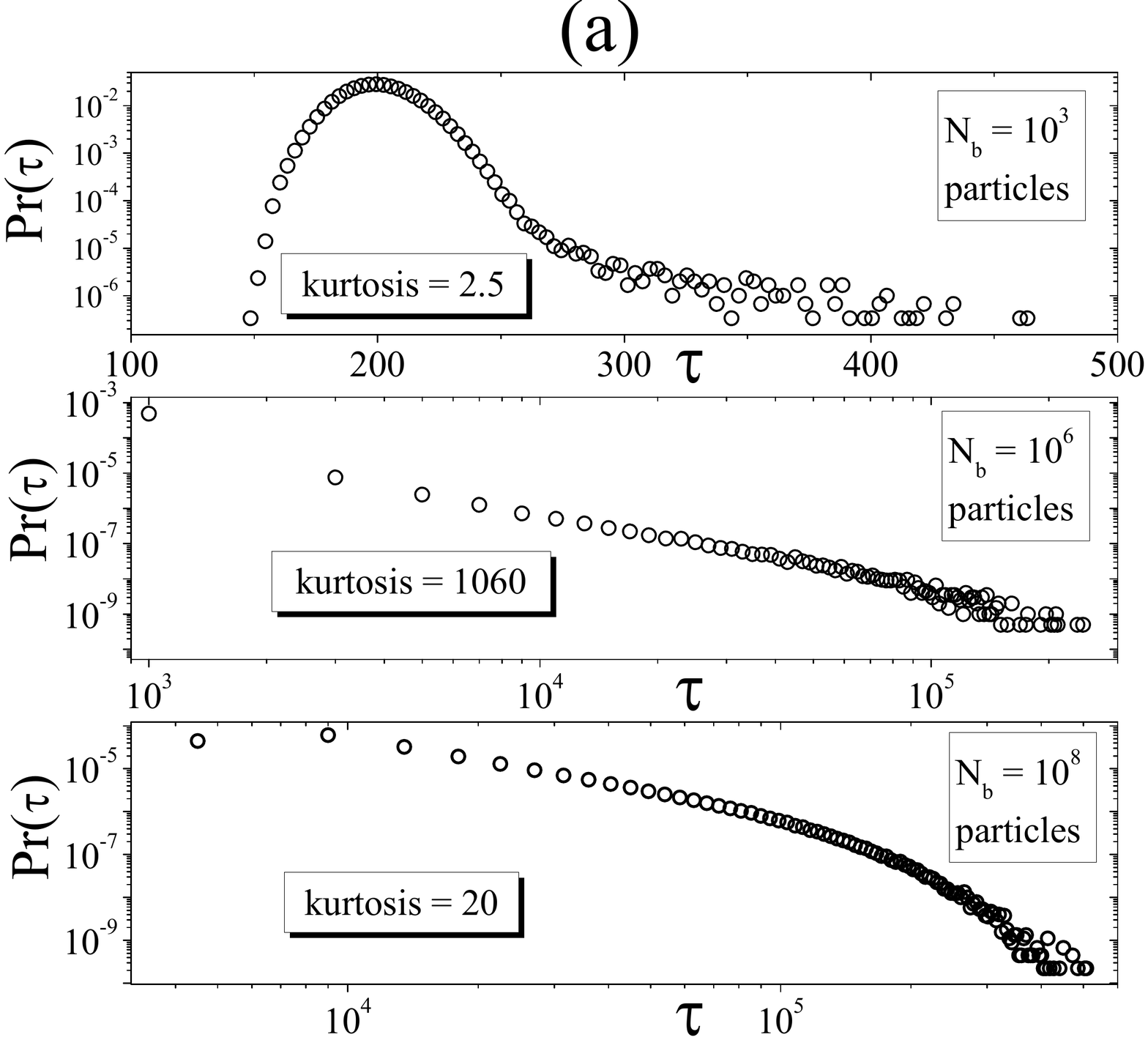} %
\includegraphics[width=0.5\columnwidth]{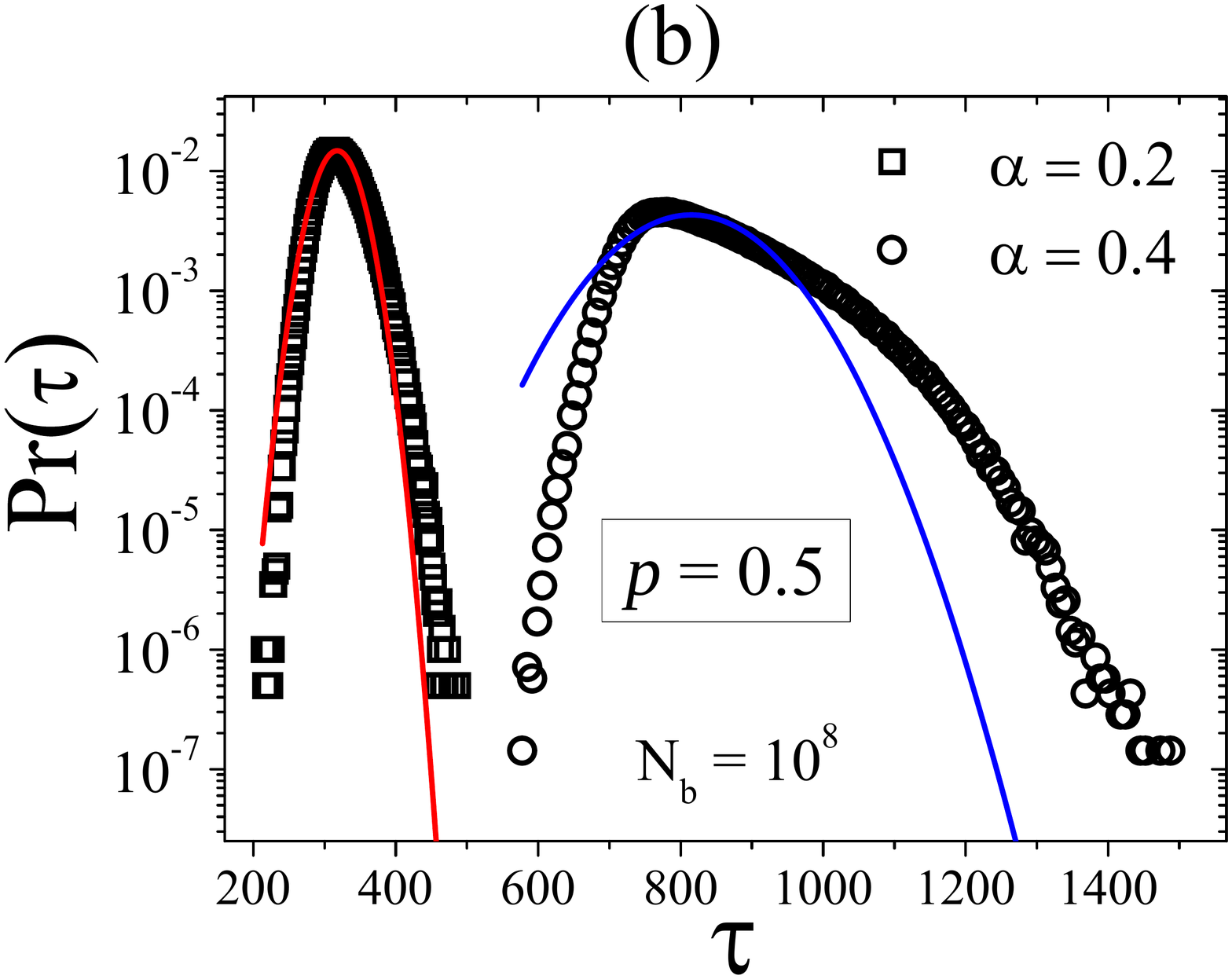}%
\includegraphics[width=0.5\columnwidth]{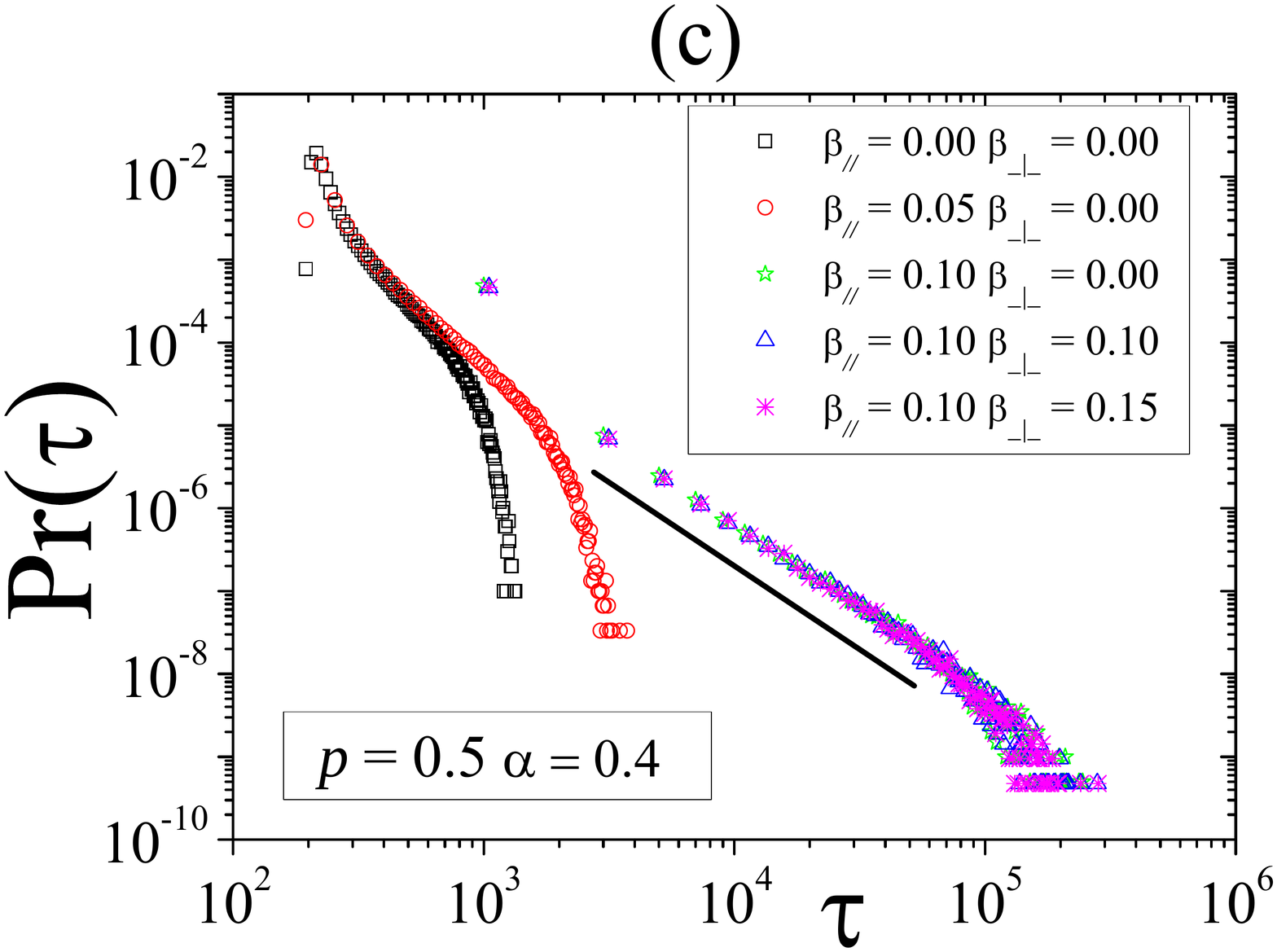}
\end{center}
\caption{MC simulations for the crossing time distribution. In all tested
situations, we used $p=0.5$ and $N_{A}=10^{6}$ particles. Plot (a) shows the
effects of the concentration of particles of the species $B$. From the top
to the bottom, we have used $N_{B}=10^{3}$, $10^{6}$, and $10^{8}$,
respectively. In these three situations, we kept $\protect\alpha =0.4$, $%
\protect\beta _{\parallel }=0.1$, and $\protect\beta _{\perp }=0$. Plot (b)
shows the effects of different values of $\protect\alpha $. Here we also
kept $N_{B}=10^{8}$ and $\protect\beta _{\parallel }=\protect\beta _{\perp
}=0$. Plot (c) shows two effects obtained by considering different values of 
$\protect\beta _{\parallel }$ for the same $\protect\beta _{\perp }$ and
different values of $\protect\beta _{\perp }$ for the same $\protect\beta %
_{\parallel }$. }
\label{Fig:crossing_times}
\end{figure}

\begin{figure}[tbp]
\begin{center}
\includegraphics[width=0.8\columnwidth]{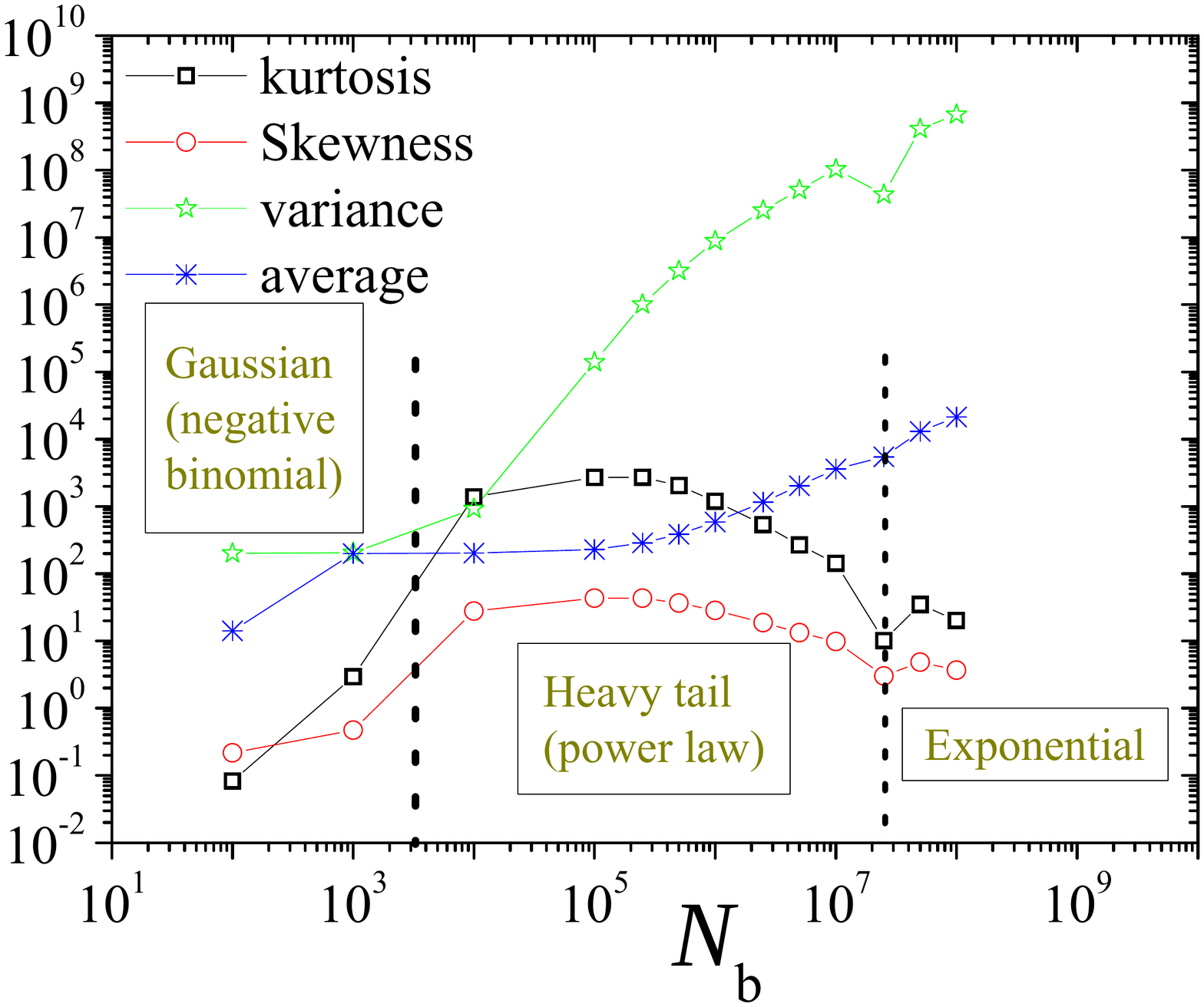}
\end{center}
\caption{Average, variance, skewness, and kurtosis of the crossing time
distribution as function of $N_{B}$. By looking at kurtosis (weight tail) of
distribution, we can observe a crossover: Gaussian behavior $\rightarrow $
power law behavior $\rightarrow $ exponential behavior.}
\label{Fig:Transition}
\end{figure}

Some stylized facts can be observed in Fig. \ref{Fig:crossing_times}, which
shows results from MC simulations for the crossing time distribution. In all
situations, we used $p=0.5$ and $N_{A}=10^{6}$ particles. Figure \ref%
{Fig:crossing_times} (a) shows the effects of different concentrations of
particles of the species $B$. From the top to the bottom, we present the
plots with $N_{B}=10^{3}$, $10^{6}$, and $10^{8}$, respectively. In these
three situations, we used $\alpha =0.4$, $\beta _{\parallel }=0.1$ and $%
\beta _{\perp }=0$. We measured the kurtosis of distribution for each case,
calculated as $\left\langle \left( \frac{\tau -\left\langle \tau
\right\rangle }{\delta }\right) ^{4}\right\rangle $which measures the weight
of tail of crossing time distribution. Here $\left\langle \tau
^{k}\right\rangle =\sum_{\tau }\tau ^{k}f(\tau )/$ $\sum f(\tau )$, where $%
f(\tau )$ is the frequency count of crossing times. Here $\delta =$ $\sqrt{%
\left\langle \tau ^{2}\right\rangle -\left\langle \tau \right\rangle ^{2}}$
is the standard deviation of crossing time distribution.

Figure \ref{Fig:crossing_times} (b) shows the effects when different values
of $\alpha$ are considered. Here, we kept $N_{B}=10^{8}$ and $%
\beta_{\parallel}=\beta_{\perp}=0$. We can observe that for $\alpha =0.4$
the deviation of a Gaussian fit (continuous curve) is easily observed.
Finally, our the plot (c) in Fig. \ref{Fig:crossing_times} shows two
different effects: different values of $\beta_{\parallel}$ for the same $%
\beta_{\perp }$ and different values of $\beta_{\perp }$ for the same $%
\beta_{\parallel}$. It is important to observe that for fixed obstacles, we
cannot observe a strong influence of $\beta_{\perp }$ on the crossing times,
i.e., the lateral motion is not important since the particles that move
laterally (changing lanes) end up encountering other obstacles whereas they
are uniformely distributed in the lattice. Different results are obtained
when one takes into account an lattice without obstacles, i.e., when both
species of particles are able to move in opposite directions.

Let us return to Fig. \ref{Fig:crossing_times} (a) in order to look into
some important details. There is an interesting crossover for the crossing
time distribution as function of $N_{B}$. We have a small kurtosis for the
top plot ($\sim 2.5$), a large kurtosis for the middle plot ($\sim 1060$),
and an intermediate kurtosis for the bottom plot ($\sim 20$). Although it is
very interesting, this transition has a simple explanation: for low values
of $N_{B}$, one have, in first approximation, a negative binomial
distribution (a Gaussian distribution at the limit) with a scattering in the
tail since there is no influence of obstacles and the particles have to
perform $L_{x}$ movements with $L_{x}/p$ trials until complete the path.
When one have an intermediate value of $N_{B}$, an interesting phenomena
arises: although there is a more probable crossing time, torn particles are
blocked by obstacles since our model is based on relative density and
therefore, clusters composed by particles of the same species have a better
confront against the particles of the other species. These torn particles
correspond to Levy flights since they have huge crossing times when compared
with the other ones. Finally, when $N_{B}$ is very large, all particles have
a hard task to cross the corridor and, differently from intermediate
situation, there is no very different crossing times leading to an
exponential behavior.

These crossovers among Gaussian, power law, and exponential behaviors are
shown in Fig. \ref{Fig:Transition} through the analysis of the average,
variance, kurtosis, and skewness as function of $N_{B}$. As it can be seen,
while the average and variance have a monotonic behavior, the kurtosis and
skewness have at first an increase and then they decrease corroborating the
crossover.

\section{Results II: species B in motion - Annealed scenario}

\label{SecResults:II}

Now, we consider that the particles of the species $B$ are also able to
move. Firstly, we analyze the fluctuations of spatial particle distribution
by considering two opposite streams of particles starting as two stripes at
the ends of the corridor (next subsection). Right after, in the second
subsection we consider the two different species randomly distributed and
mixed in the corridor and we analyze the fluctuations on the distillation
time, i.e., the needed time to separate the two species.

\subsection{Properties of the spatial particle distribution}

As we have already explored the effects of crossing times of particles of
the species $A$ when particles of the species $B$ work as fixed obstacles,
now we would like to study some effects when the particles of the species $B$
are able to move in the lattice. For this purpose, we analyze the properties
of a scenario where two streams of different species of particles interact
with each other when moving in opposite direction. So, we consider the
condition previously set up to the particles of the species $A$ as valid for
both species. The condition for the species $B$ is then given by: 
\begin{equation*}
n_{B}(j,k;t=0)=\left\{ 
\begin{array}{lll}
\frac{N_{B}}{L_{y}} & \text{if} & 
\begin{array}{l}
j=L_x, \\ 
k=1,\cdots,L_{y}%
\end{array}
\\ 
&  &  \\ 
0 & \text{if} & 
\begin{array}{l}
1<j<L_{x}-1, \\ 
k=1,\cdots,L_{y}%
\end{array}%
\end{array}%
\right.
\end{equation*}

\begin{figure}[htbp]
\begin{center}
\includegraphics[width=0.85\columnwidth]{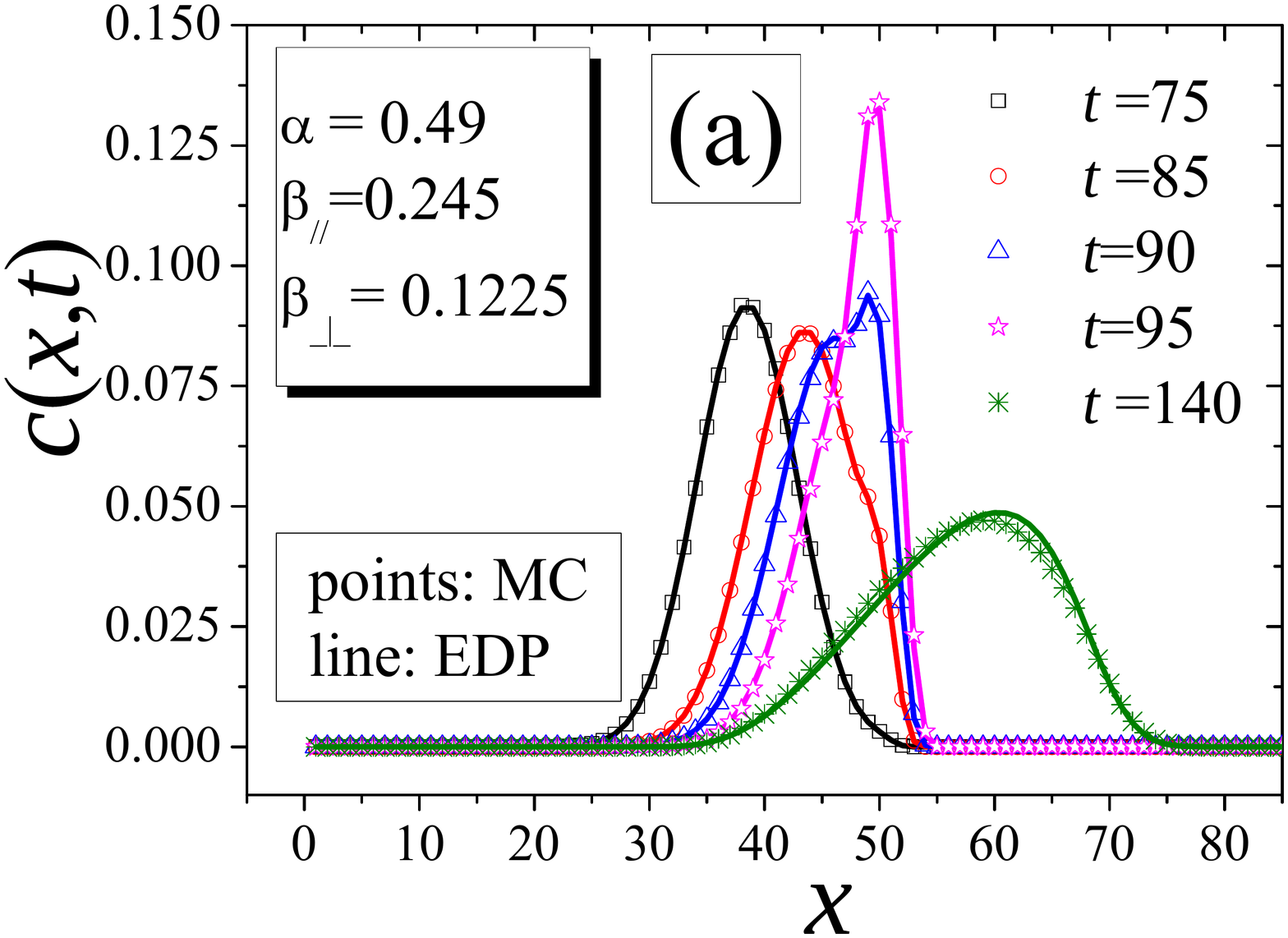} %
\includegraphics[width=0.85\columnwidth]{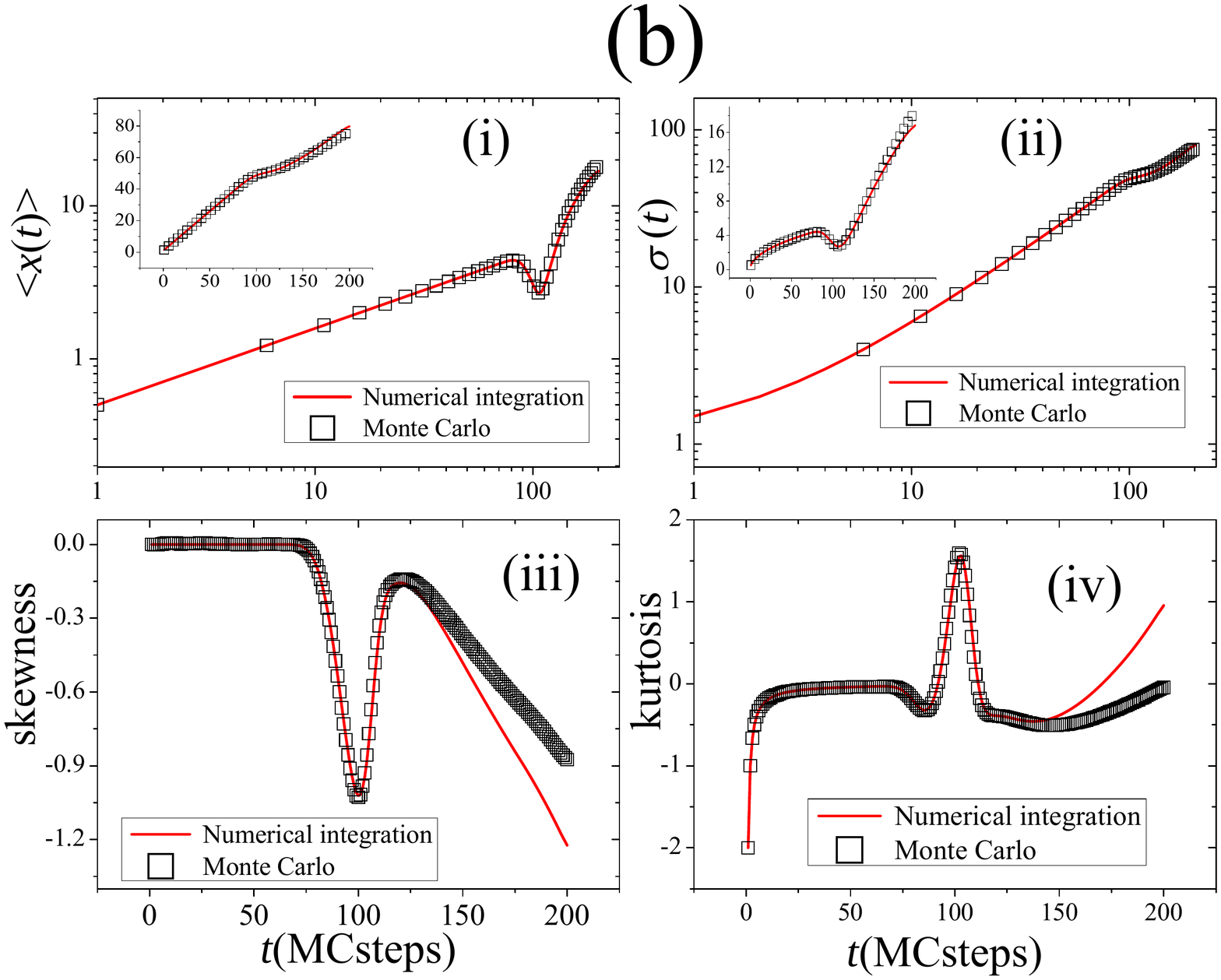}
\end{center}
\caption{\textbf{Plot (a)}: Density of particles $A$ ($B$ is symmetric since
the initial conditions are similar) in different times of evolution. \textbf{%
Plot (b)}: Time evolution of fluctuations of the density of particles: 
\textbf{average, standard deviation, skewness and kurtosis}.}
\label{Fig:Properties_of_concentration}
\end{figure}

Here, we are interested in the study of the scenario when the two species
meet each other and interact by following the dynamics previously defined in
this work. After starting the evolution of the two stripes, we follow the
time evolution of the shape of the density of particles for several times as
can be seen in Fig. \ref{Fig:Properties_of_concentration} (a) as well as of
the time evolution of the parameters which describe such density, i.e., its
fluctuations (plot (b), in the same figure).

We can observe in Fig. \ref{Fig:Properties_of_concentration} (a) that the
density of particles $c(x,t)$ (at cell $j=x/a$ and time $t=l\Delta $) is
deformed during the interaction of the species from $t=85$. Here it is
important to notice that a marginalization over $y$\ dimension was
conveniently performed according to our objectives. Figure \ref%
{Fig:Properties_of_concentration} (b) shows that this deformation is
captured by valleys in the skewness (iii):$\ \left[ \sum_{x}\left( \frac{%
x-\left\langle x(t)\right\rangle }{\sigma (t)}\right) ^{3}c(x,t)\right]
/\sum_{x}c(x,t)$ and peaks in the kurtosis (iv): $\left[ \sum_{x}\left( 
\frac{x-\left\langle x(t)\right\rangle }{\sigma (t)}\right) ^{4}c(x,t)\right]
/\sum_{x}c(x,t)$, with $\left\langle x(t)\right\rangle =\left[
\sum_{x}xc(x,t)\right] /\sum_{x}c(x,t)$ and $\sigma (t)=\left[ \left\langle
x^{2}(t)\right\rangle -\left\langle x(t)\right\rangle ^{2}\right] ^{1/2}$,
that increase in absolute value during the evolution, leaving the Gaussian
behavior (represented by the distribution for $t=75$ as showed in plot (a))
to become a heavy tail distribution. We also can observe that such anomalous
behavior is also captured by the average position $\left\langle
x(t)\right\rangle $, in the plot (i) as well as by the standard deviation of
the position $\sigma (t)$ of the particles according to plot (ii). In both
plots we use $\alpha =0.49$, $\beta _{\parallel }=0.245$ and $\beta _{\perp
}=0.1225$. In Fig. \ref{Fig:Properties_of_concentration} (a) and (b) we also
present the comparison between the results obtained through MC simulations
and EDP solutions. As shown, they are in good agreement to each other.

\subsection{Distillation time}

\begin{figure}[ht]
\begin{center}
\includegraphics[width=0.8\columnwidth]{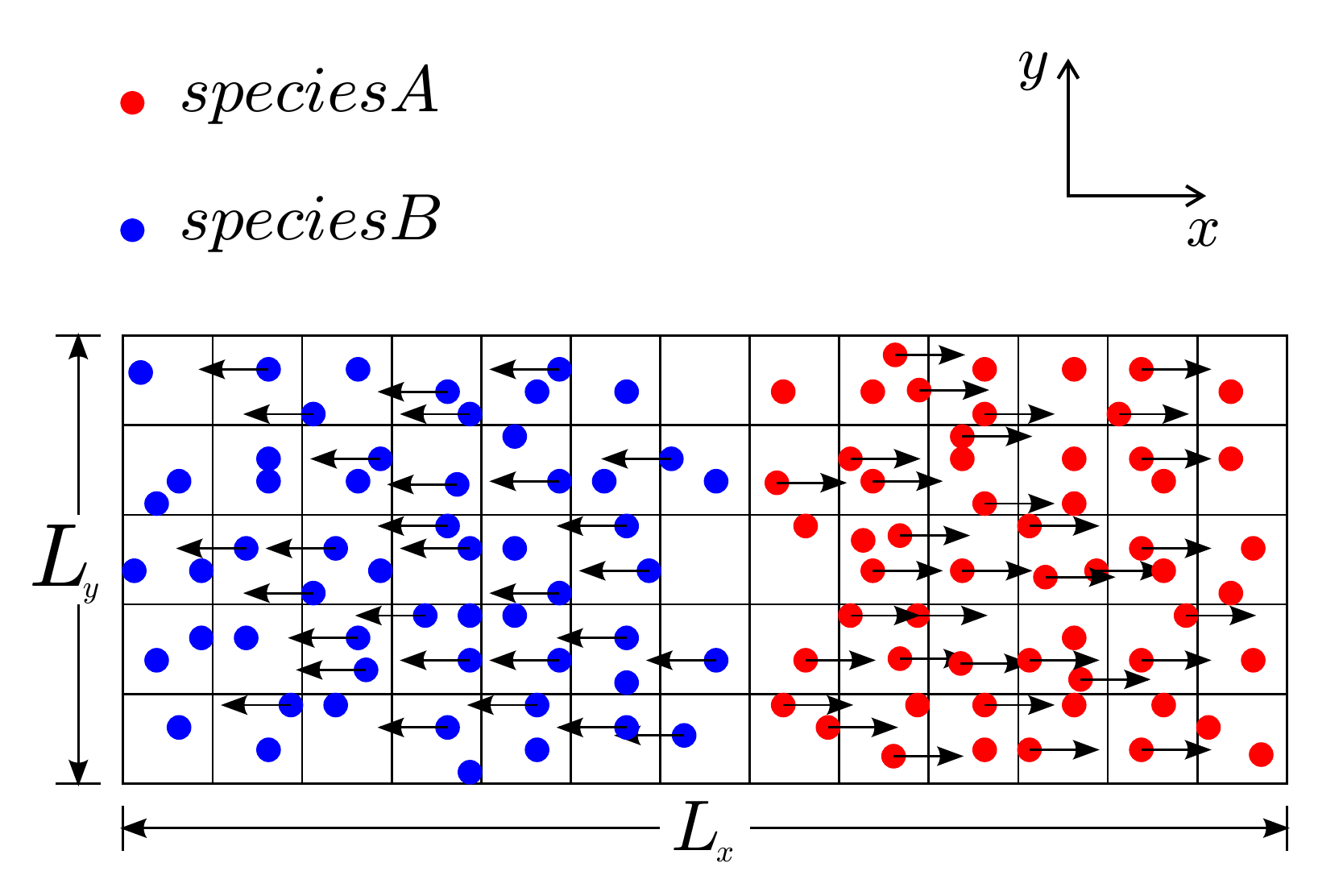}
\end{center}
\caption{Illustration of a complete distillation between the species $A$ and 
$B$. This means that there is a vertical line separating the two species in
the corridor.}
\label{Fig.complete_destilation}
\end{figure}

As the two species of particles follow opposite directions, an interesting
situation can be observed when we mix them in the corridor at the initial
stage of the evolution and analyze the needed time to a complete separation
of the two species (distillation). In Fig. \ref{Fig.complete_destilation} we
show a typical situation in which the species are completed distillate.

\begin{figure*}[htbp]
\begin{center}
\includegraphics[width=0.5%
\columnwidth]{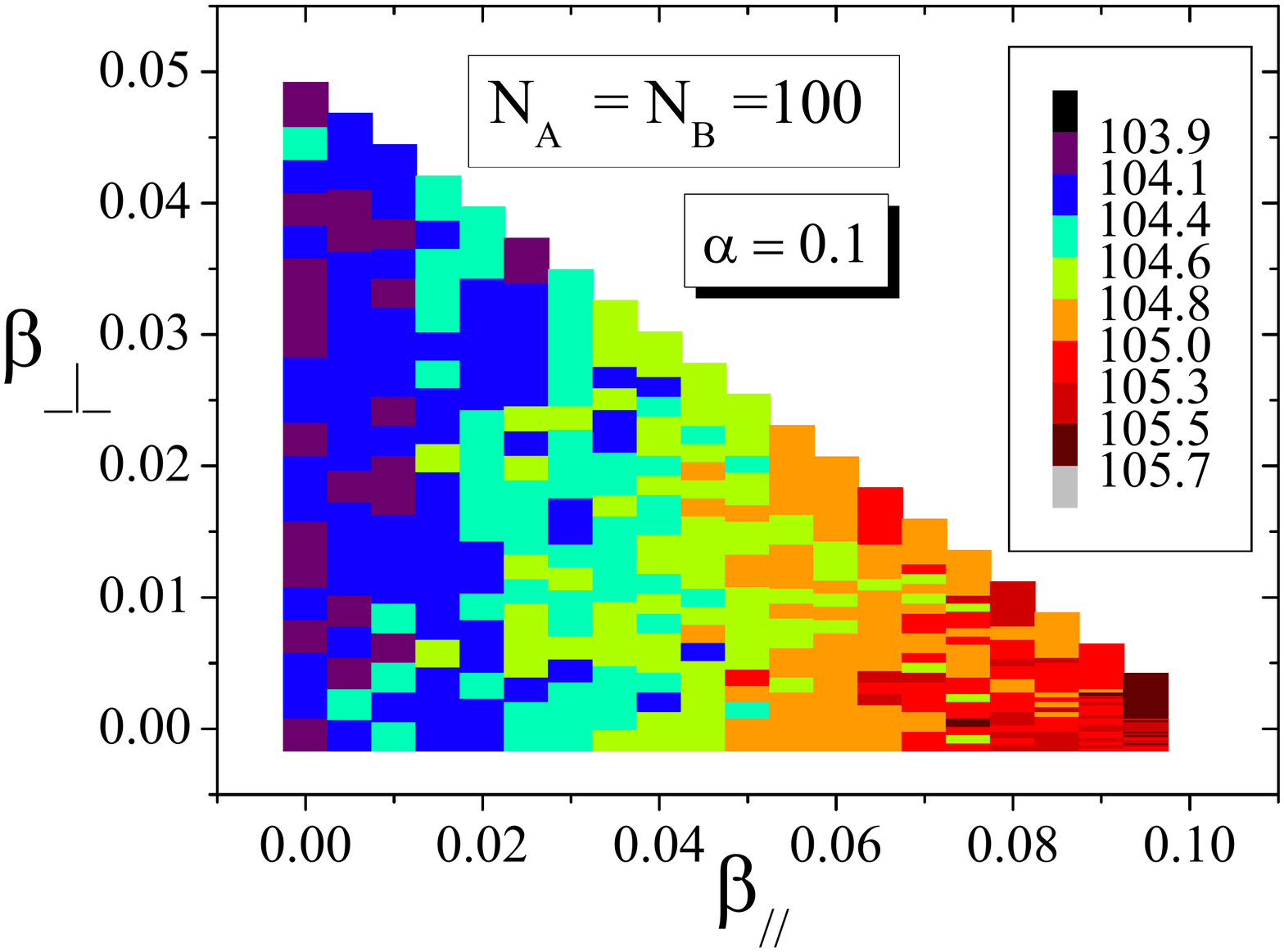}%
\includegraphics[width=0.5%
\columnwidth]{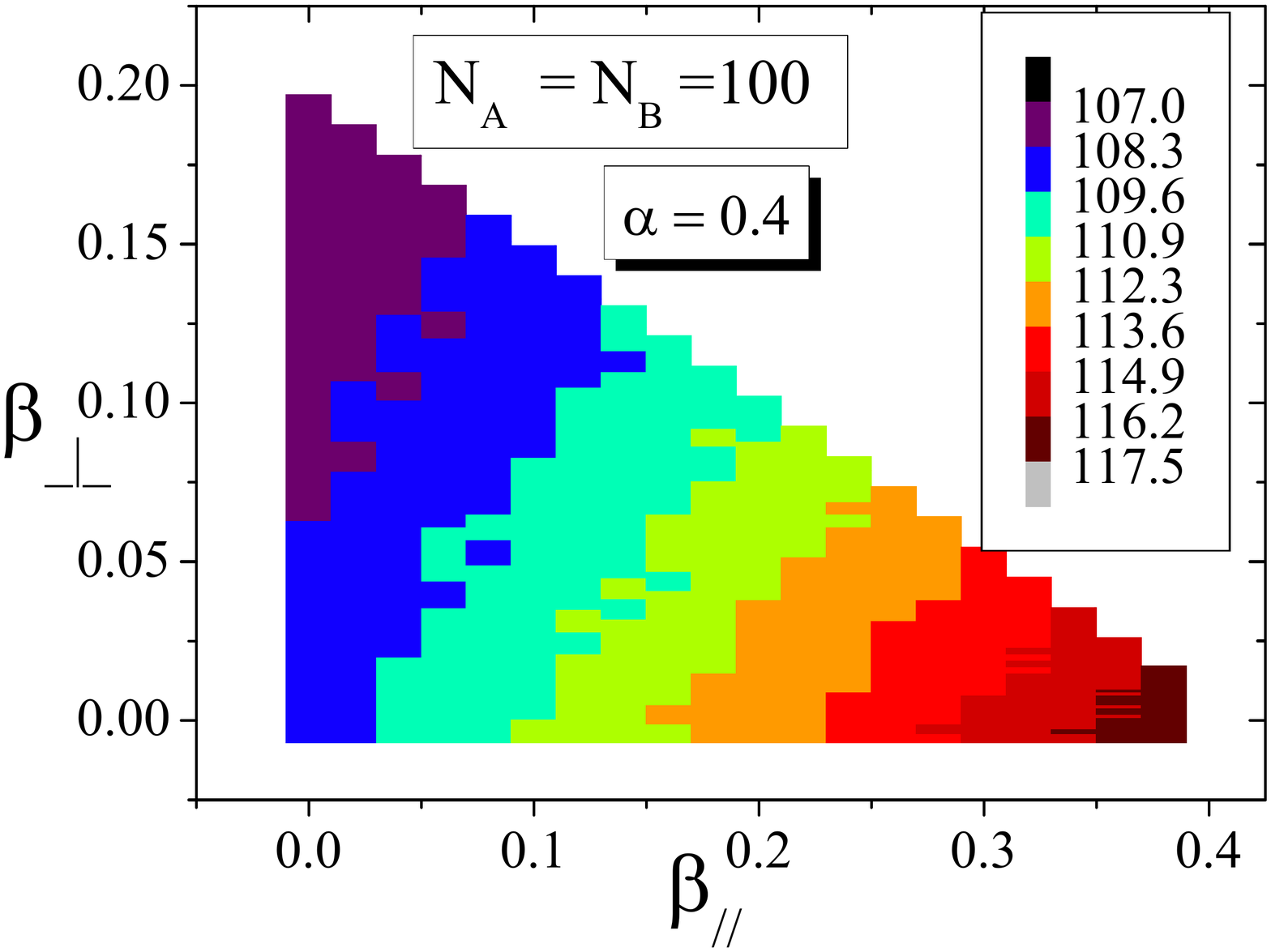} %
\includegraphics[width=0.5%
\columnwidth]{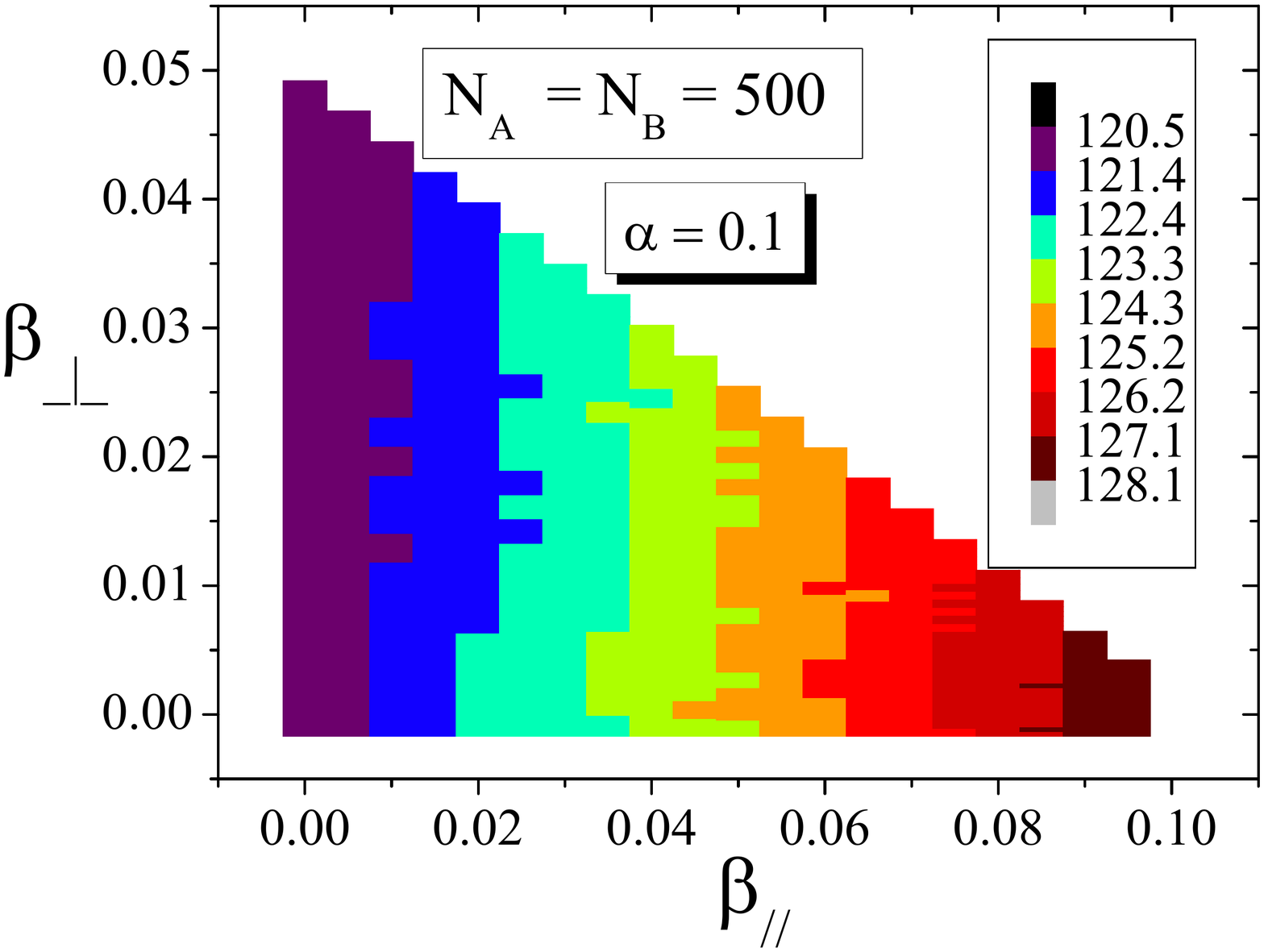}%
\includegraphics[width=0.5%
\columnwidth]{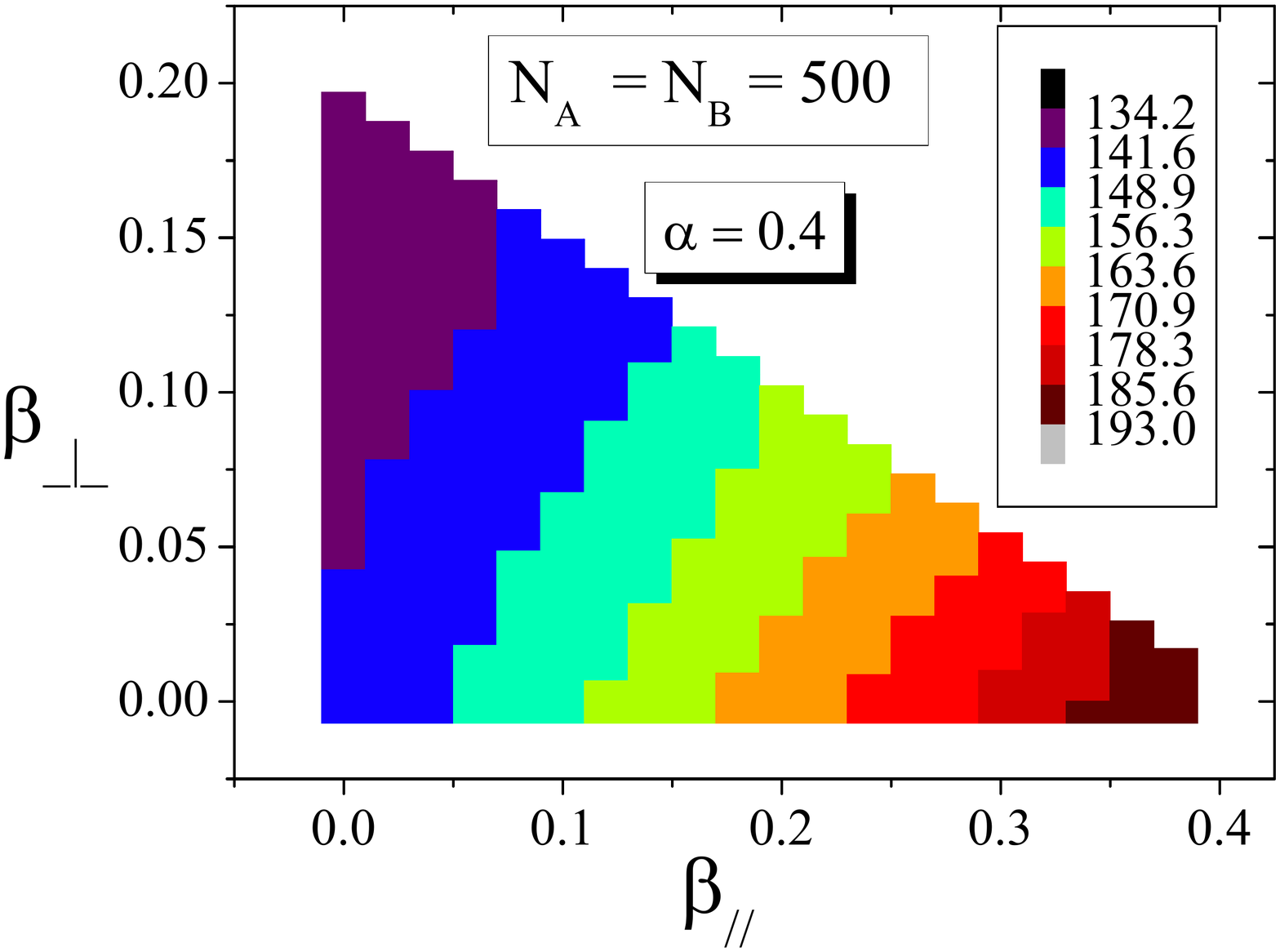} %
\includegraphics[width=0.5%
\columnwidth]{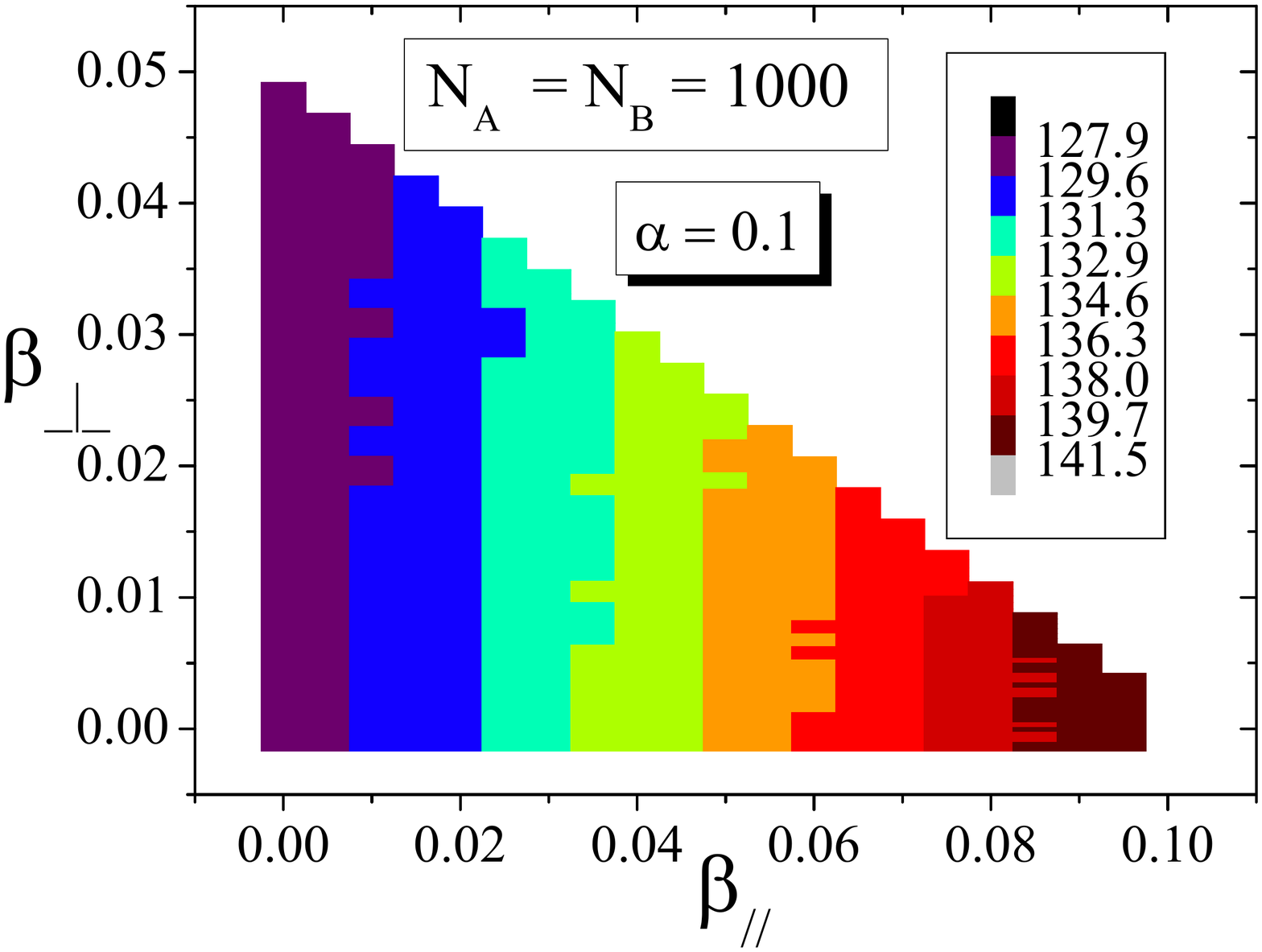}%
\includegraphics[width=0.5%
\columnwidth]{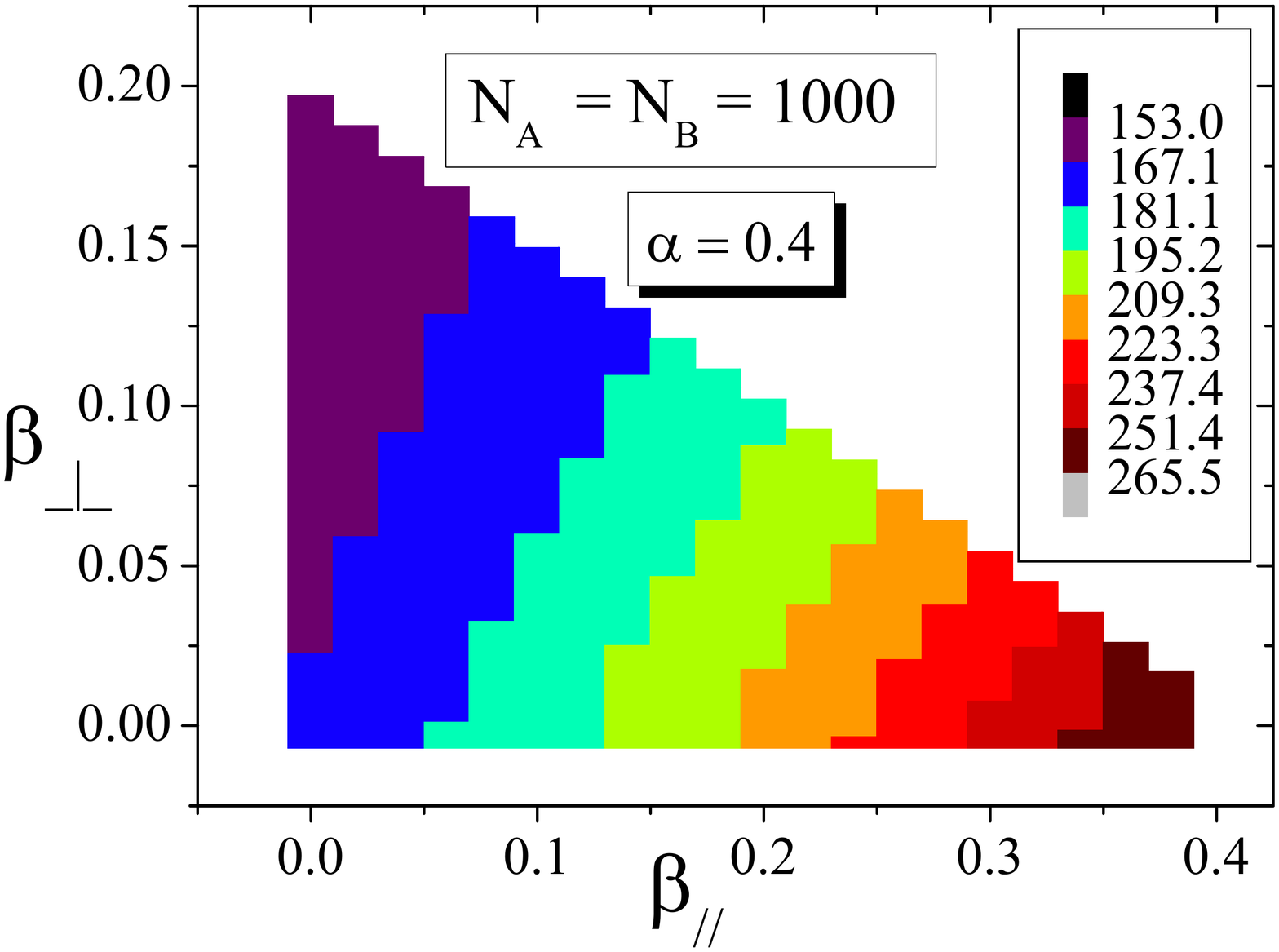}
\end{center}
\caption{Color map of the distillation time for different values of $\protect%
\beta_{\parallel}$ and $\protect\beta_{\perp}$ considering different
concentrations of particles for two different values of $\protect\alpha$.}
\label{Fig.Color_Map_destilation_time}
\end{figure*}

The needed time for the system to achieve the complete separation of the
species is computed through numerical MC simulations for all possible pairs $%
\left(\beta_{\parallel},\beta_{\perp}\right)$ for two different values of $%
\alpha$, one low: 0.1, and another high: 0.4. This procedure was repeated
considering three different number of particles $N=N_{A}=N_{B}=100$, 500,
and 1000. Here, the main idea is to check if $\beta_{\perp}$ can affect the
distillation time. Figure \ref{Fig.Color_Map_destilation_time} shows that
for $\alpha=0.1$ the effects of $\beta_{\perp }$ on the distillation time
are not perceptible, however $\alpha=0.4$ leads to a reduction of the
distillation time. 

This reduction is more relevant when $\beta_{\parallel}$ is higher, i.e.,
the movements to the sides are important when the combination of the
resistance factor ($\alpha$) and the frontal collision effects ($%
\beta_{\parallel}$) leads to a real stagnation in the movement ability.
Here, it is interesting to monitor the order parameters $\Phi_{cell}(t)$ and 
$\Phi_{\perp }(t)$ as function of time during the distillation previously
described in Fig. \ref{Fig.Color_Map_destilation_time}. We hope that when $%
t\rightarrow \infty $, both $\Phi_{cell}$ and $\Phi_{\perp }\rightarrow 1$
since the system is completely distillate.

\begin{figure}[htbp]
\begin{center}
\includegraphics[width=%
\columnwidth]{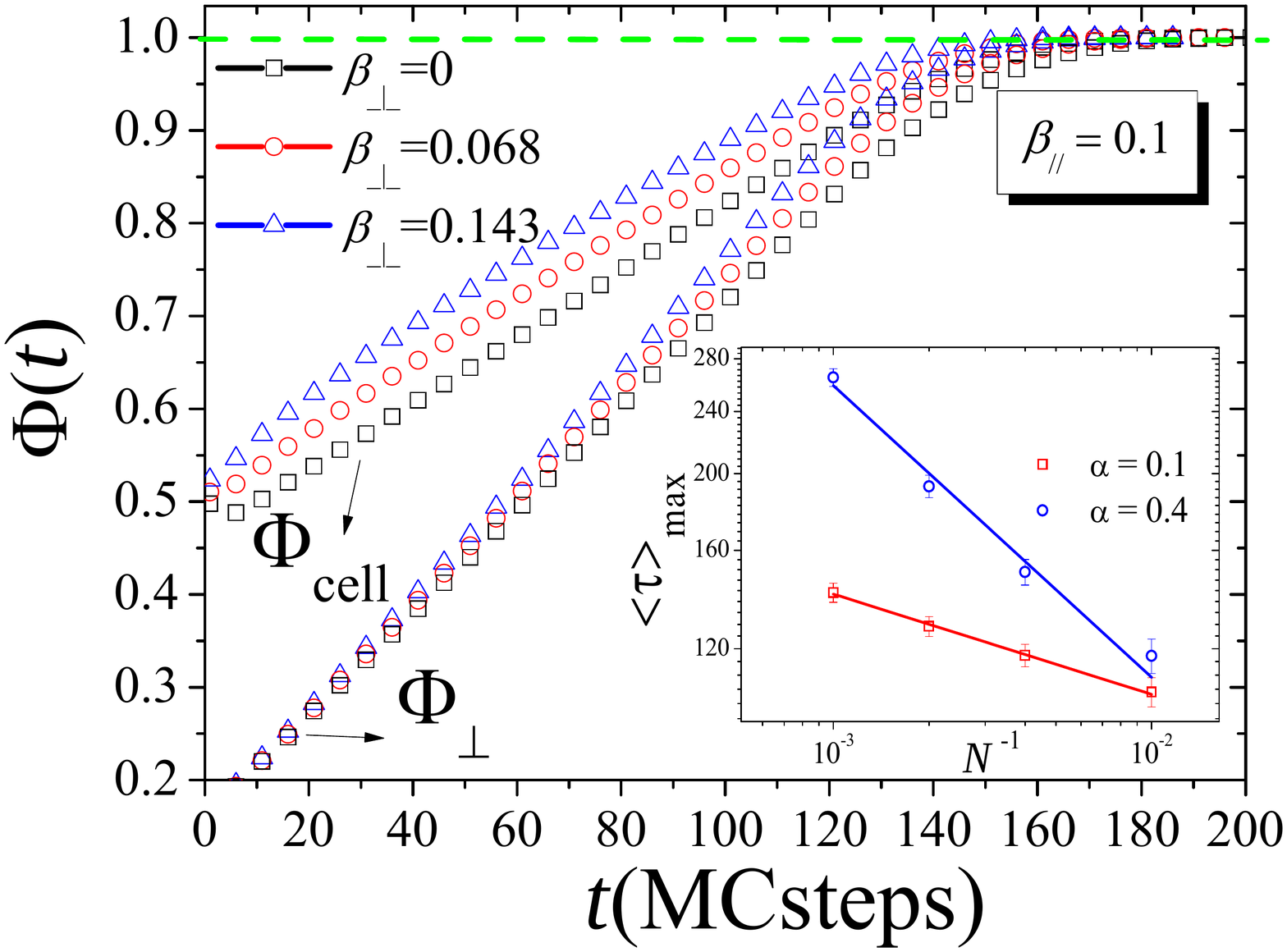}
\end{center}
\caption{Time evolution of the order parameters $\Phi_{cell}$ and $%
\Phi_{\perp}$ for $\protect\beta_{\parallel}$ fixed and for different values
of $\protect\beta_{\perp}$. We can observe that both parameters converge to
1 (indicating that the system is completely distillate), and the segregation
is greater as $\protect\beta_{\perp}$ increases. The inset plot
shows as the distillation times scales with the number of particles.}
\label{Fig:Destilation_order_parameters}
\end{figure}

In Fig. \ref{Fig:Destilation_order_parameters} we show that both parameters
converge to 1 (indicating that the system is completely distillate), and the
segregation is greater as $\beta _{\perp }$ increases. A simple analysis of
the finite size scaling effects of the average time of the distillation time
is shown in the inset of this figure. This plot is obtained for the worst
case ($\beta_{\perp }=0$ and the maximal allowed $\beta_{\parallel}$, i.e., $%
\beta_{\parallel}=\alpha $) as function of the inverse number of particles
in the system: $N^{-1}=N_{A}^{-1}=N_{B}^{-1}$, and is presented in log-log
scale for the two studied values of $\alpha $.

Whereas we have already explored the properties of the distillation time and
the dependence of the different factors that change this quantity, we study,
in the following section, the band formation which appears when the
particles move in a ring which is obtained by considering periodic boundary
conditions in the longitudinal direction.

\section{Results III: Periodic boundary conditions - motion in a ring}

\begin{figure}[htbp]
\begin{center}
\includegraphics[width=0.6\columnwidth]{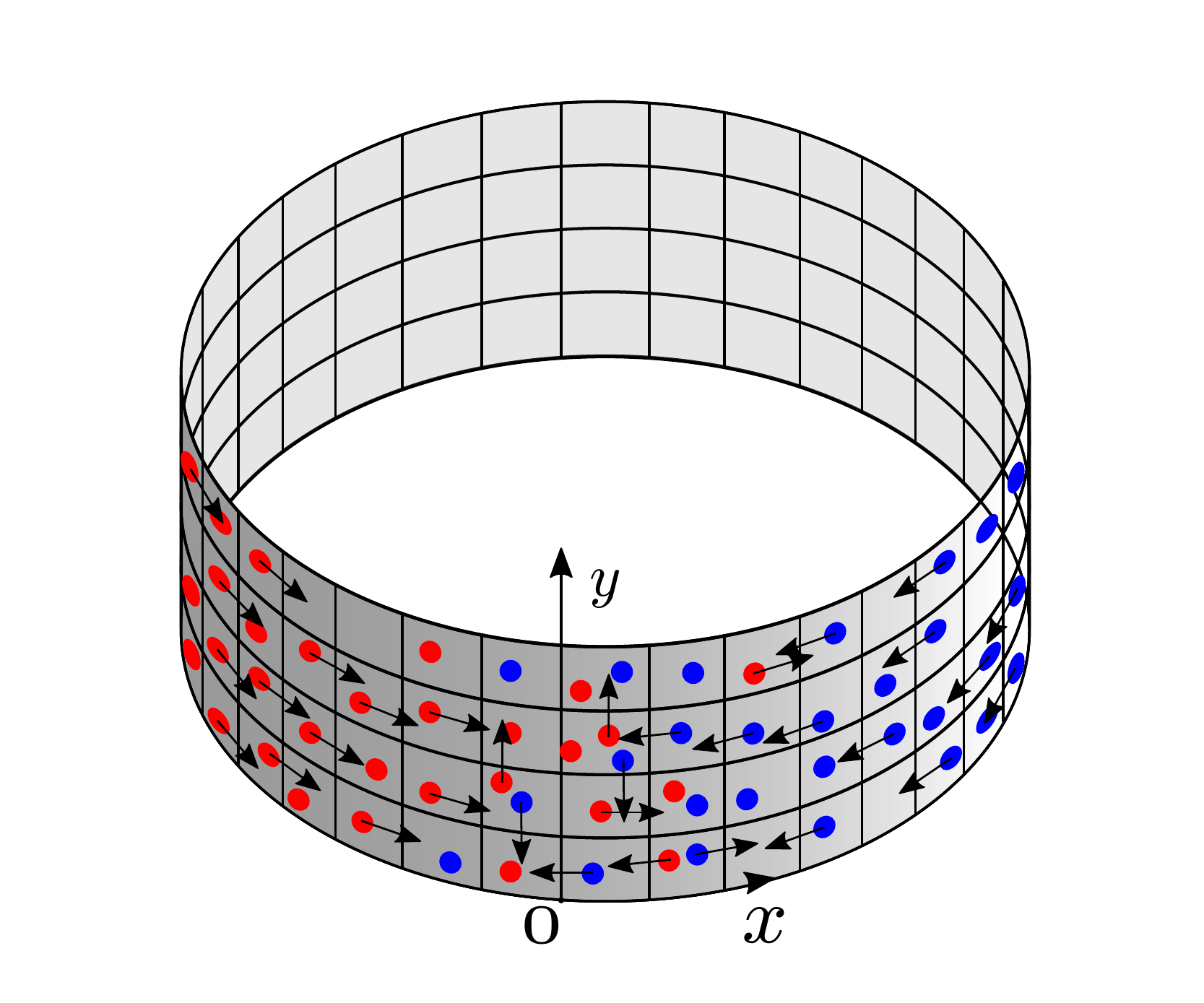}
\end{center}
\caption{Particle counterflow of different species moving in a ring.}
\label{Fig:ring}
\end{figure}

Now, we consider the two species of particles moving in a ring, in the
opposite direction to each other, i.e., one species of particles is moving
in the counterclockwise and the another one is moving in the clockwise.
Alternatively, this situation can be thought as if particles could enter in
the corridor according to some rate if they are of the same species that
those which are reaching the end of corridor (ring topology). The idea is
represented in Fig. \ref{Fig:ring}.

First of all, we perform MC simulations considering that the species have
the same number of particles ($N_{A}=N_{B}=10^{5}$) and are uniformily
distributed in the corridor. Thus, we monitor the concentrations of
particles in the corridor as the time evolves, for a typical set of
parameters $\alpha =0.45$ and $\beta _{\parallel}=\beta_{\perp }=0.15$ in
order to observe the process of relaxation toward the formation of bands. We
analyze the frames for three different MC steps: $t=1$, $1100$, and $64000$.
Band formations can be observed after a long time corroborating the ordering
of the species which move in the same direction in the stationary stage of
the evolution.

\begin{figure*}[htbp]
\begin{center}
\includegraphics[width=0.5\columnwidth]{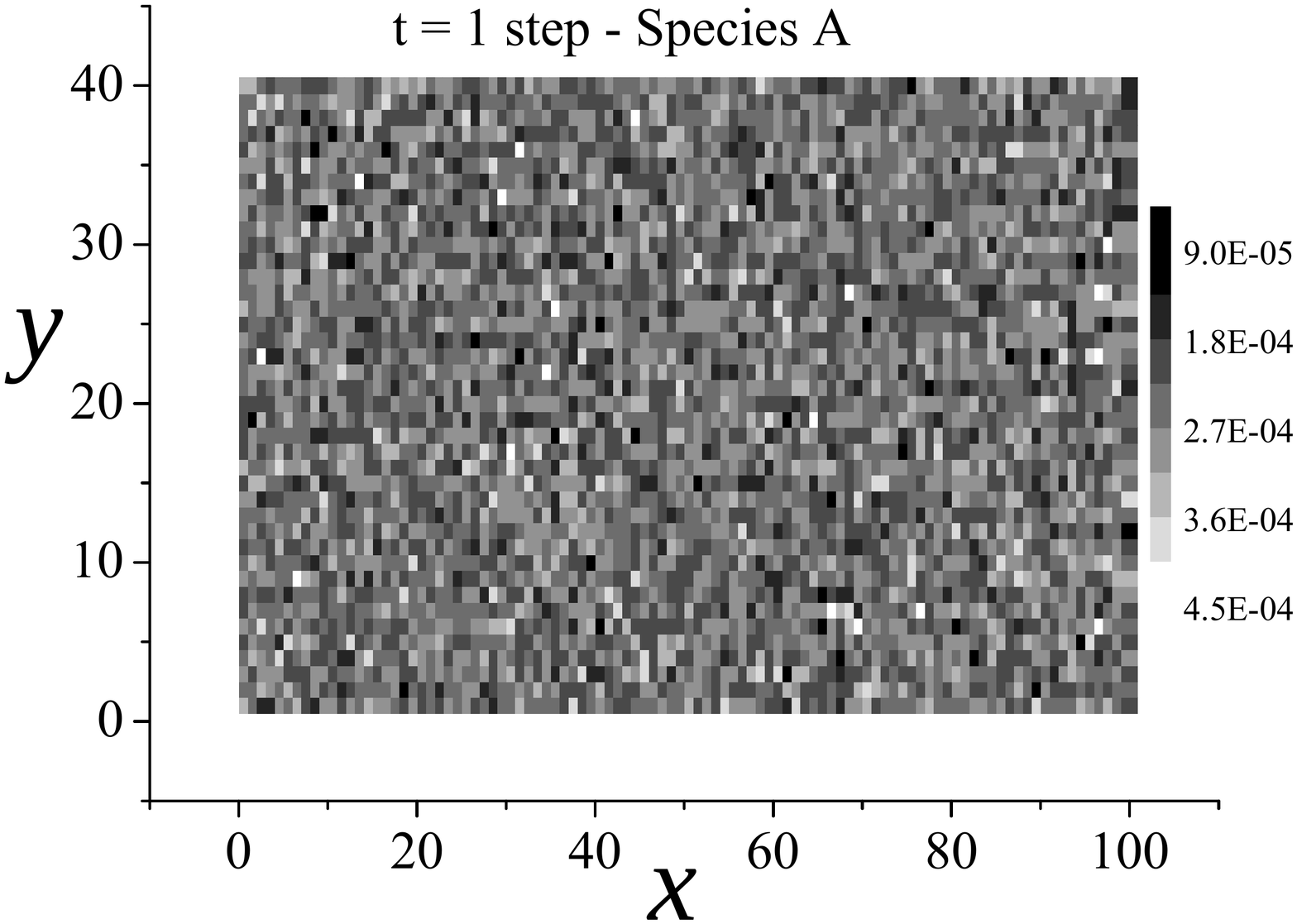}%
\includegraphics[width=0.5\columnwidth]{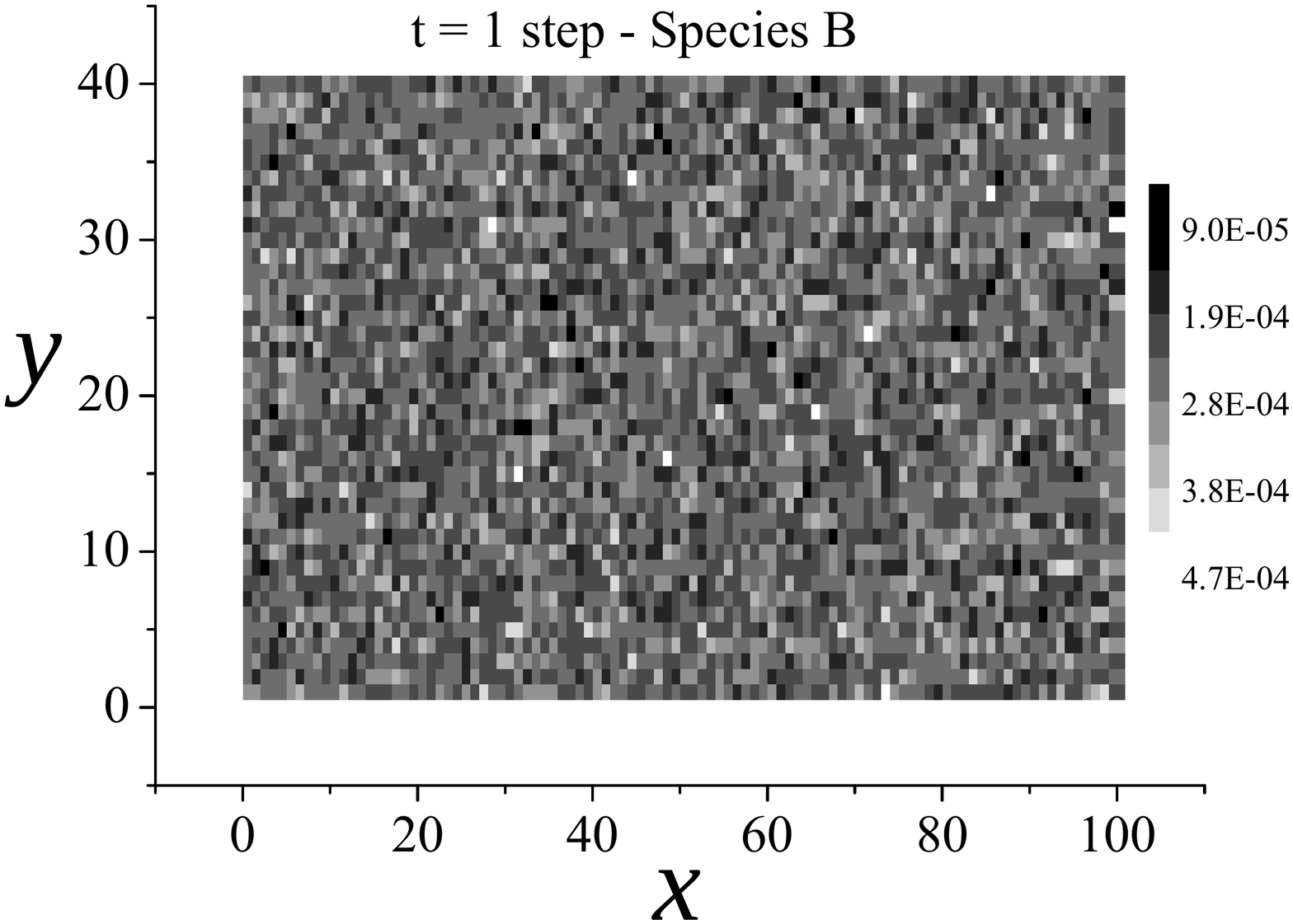} %
\includegraphics[width=0.5\columnwidth]{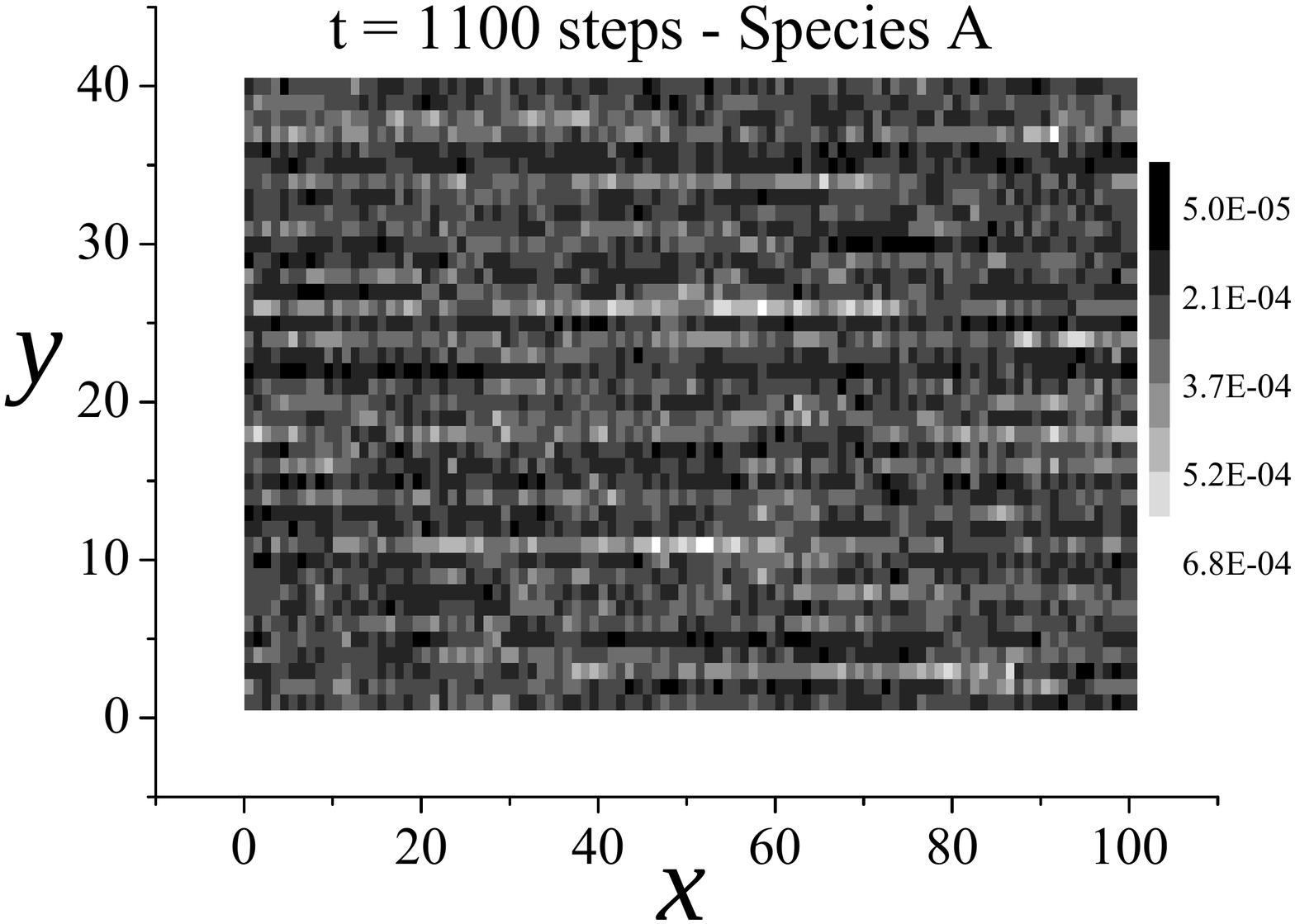}%
\includegraphics[width=0.5\columnwidth]{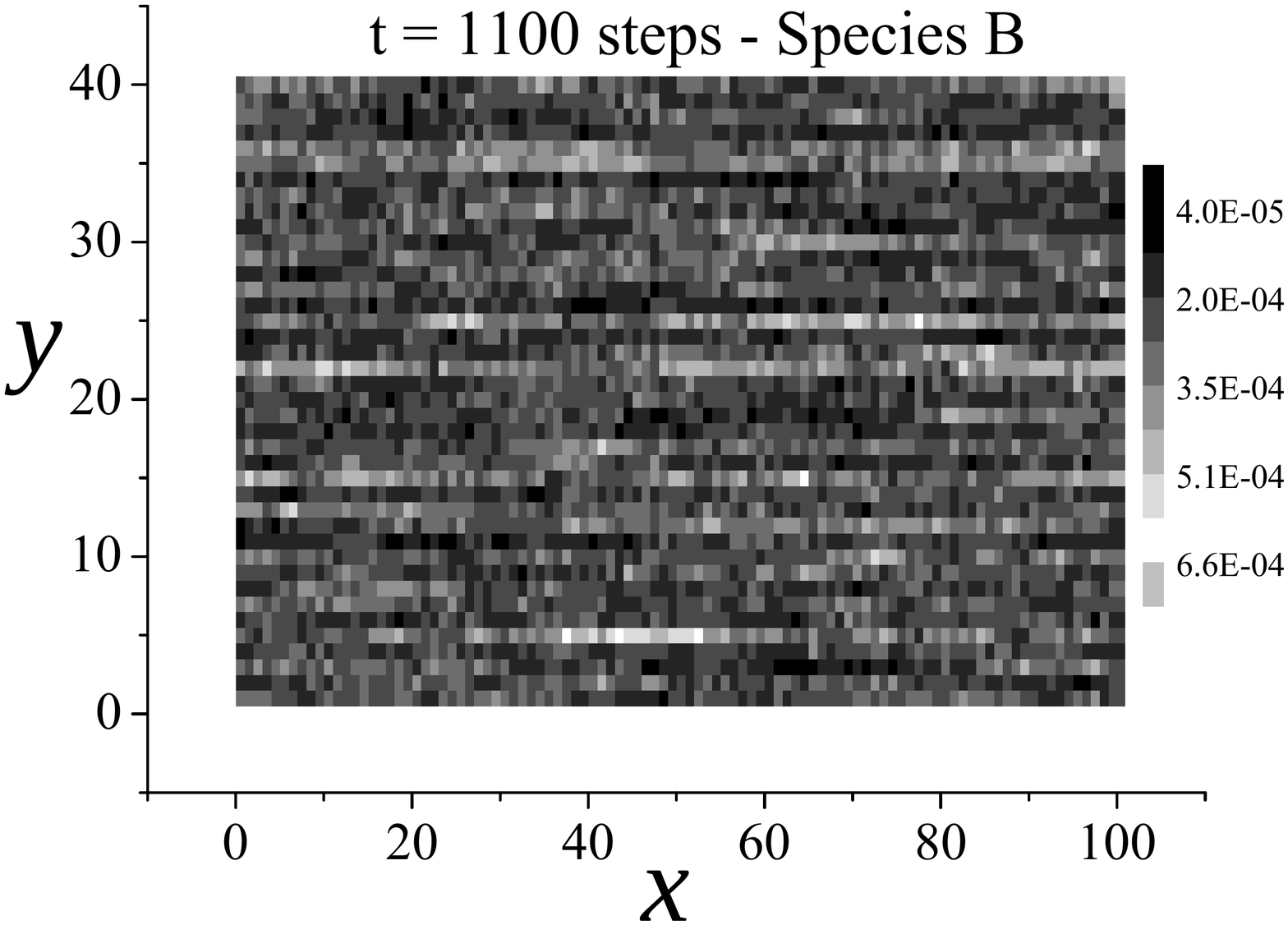} %
\includegraphics[width=0.5\columnwidth]{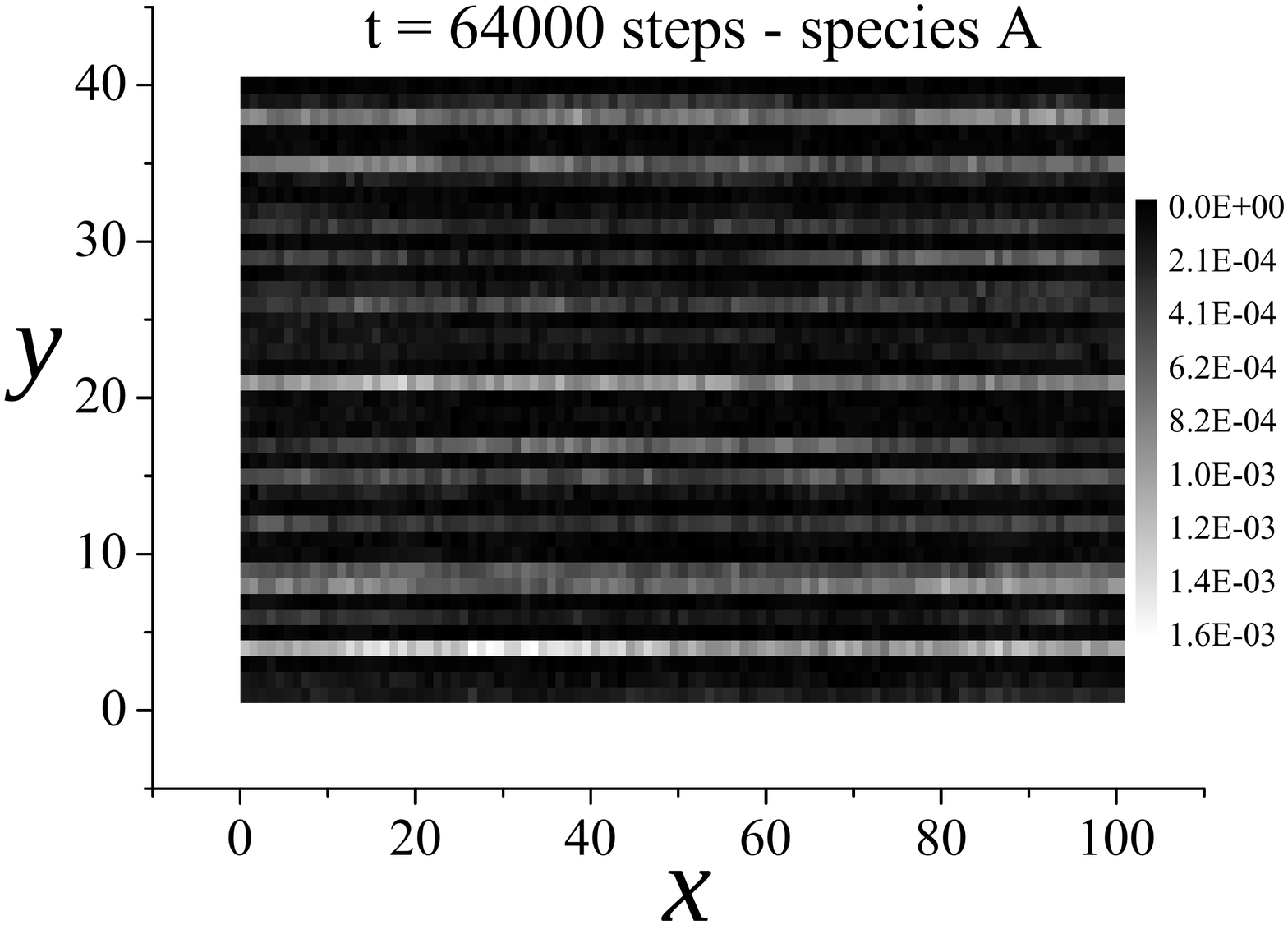}%
\includegraphics[width=0.5\columnwidth]{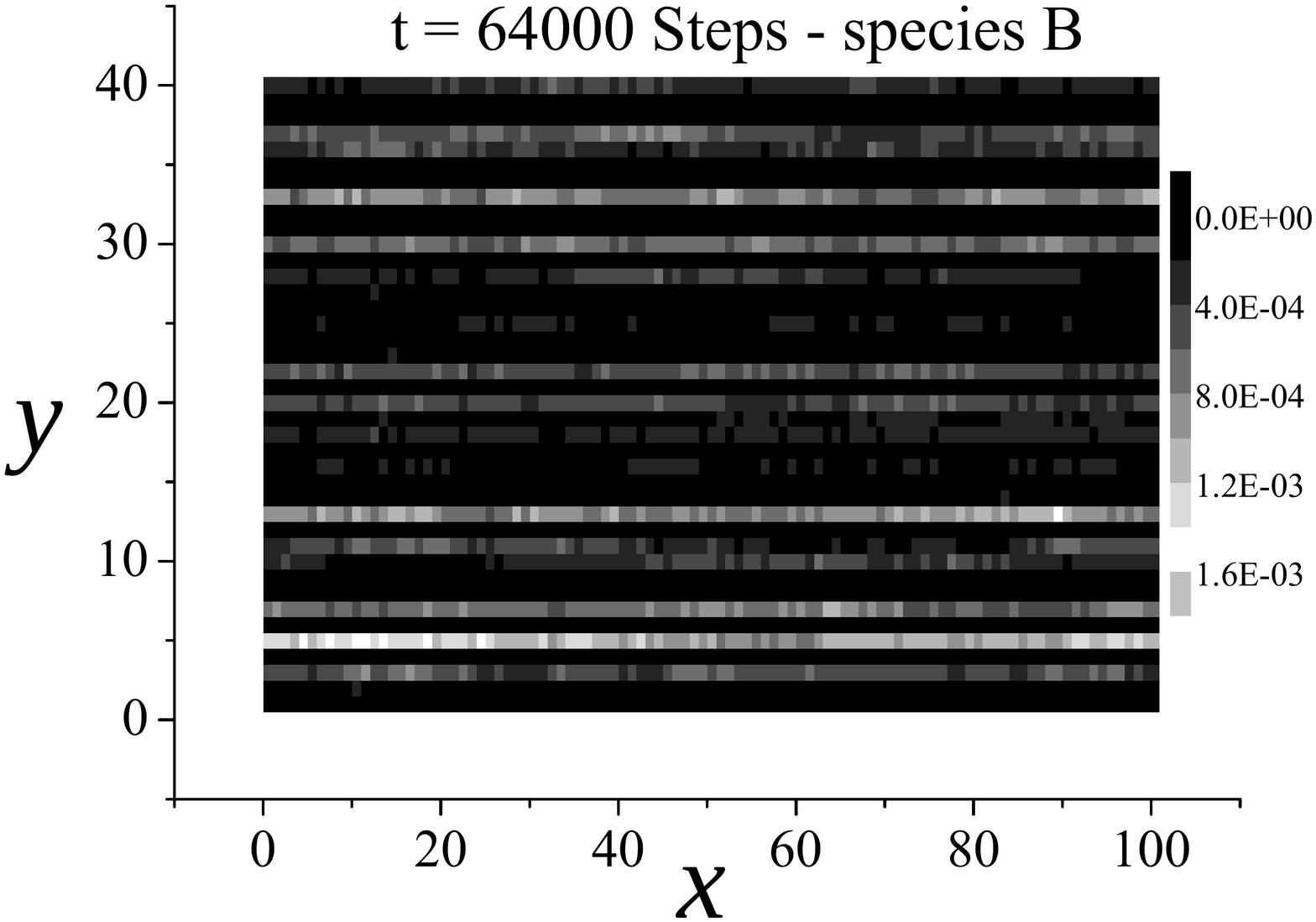}
\end{center}
\caption{Frames of the density of particles of the species $A$ (left side)
and $B$ (right side) for different instants of time when the species move in
opposite direction. The particles are uniformily distributed at the
beginning of the evolution and follow the ring topology. Band formations can
be observed in the final of evolution.}
\label{Fig:Snapshots_from_MC_homogenous}
\end{figure*}

\begin{figure}[htbp]
\begin{center}
\includegraphics[width=\columnwidth]{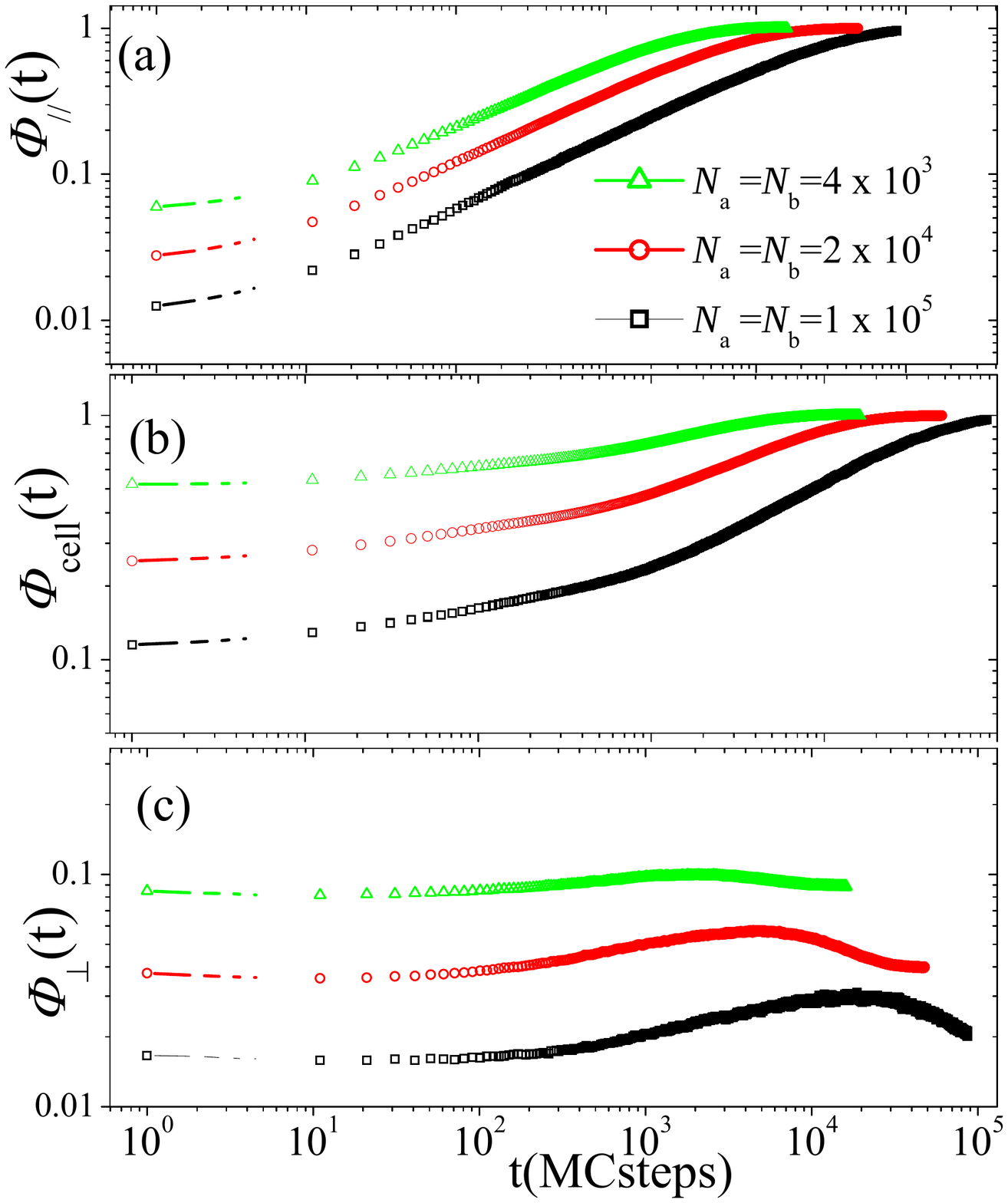}
\end{center}
\caption{Relaxation of the order parameters. We can see that $%
\Phi_{\parallel}$ and $\Phi_{cell}$ go to 1 when $t\rightarrow \infty $
corroborating the ordering of the species in longitudinal bands. However the
segregation by bands in the perpendicular direction to the field is not
observed.}
\label{Fig:order_parameters_homogenous}
\end{figure}

Such patterns can be better understood if we analyze the order parameters $%
\Phi _{\parallel}(t)$, $\Phi_{cell}(t)$, and $\Phi_{\perp}(t)$ as function
of the time. We can observe in Fig. \ref{Fig:order_parameters_homogenous}
that $\Phi_{\parallel}(t)$ goes asymptotically to 1, which corroborates the
longitudinal bands observed in Fig. \ref{Fig:Snapshots_from_MC_homogenous}.
Likewise, we can see that $\Phi_{cell}(t)$ converges to 1 since a
segregation by bands (longitudinal or vertical) implies in a segregation by
cells. However, $\Phi_{\perp}(t)$ possesses lower values showing that the
segregation by vertical bands (perpendicular to the orientation field of
particles) is not expected.

Here, it is interesting to make an analogy with some real physical systems.
Vissers et al. \cite{Vissers2011} observed an interesting lane formation in
driven mixtures of oppositely charged colloids when considering periodic
boundary conditions and their results corroborate our theoretical findings
in a purely stochastic model (see Fig. 1 in Ref. \cite{Vissers2011}). In
other publication, Vissers et al. \cite{VissersPRL2011} showed that vertical
bands appear when one apply an ac field. In a upcoming paper, we will
explore the effects due to an ac field in a stochastic model in order to
reproduce such experimental results.

In a recent work, Oliveira et al. \cite{Pinho2016} proposed a model of
pedestrian behavior and obtained two-lane ordered state emerging with
asymmetrically shaped walls. So, they were are able to organize the flow of
pedestrians moving in opposite directions by solving Langevin-like equations
with Stokesian drag force. Inspired in this work, we propose an interesting
computational experiment by using our model. However, differently from that
study, we focus our attention to initial conditions of the particles. So, we
start with the particles of the species $A$ initially concentrated in a
vertical stripe asymmetrically distributed along $y$: 
\begin{equation}
n_{A}(j,k;t=0)=\left\{ 
\begin{array}{lll}
\frac{2N_A}{L_y}(L_y-k) & \text{if} & j=1, k=1,\cdots,L_y \\ 
&  &  \\ 
0 & \text{if} & 
\begin{array}{l}
2\leq j\leq L_{x}, \\ 
k=1,\cdots,L_{y}%
\end{array}%
\end{array}
\right.  \label{Eq.assymetry}
\end{equation}
and the particles of the species $B$ are uniformly distributed on the
corridor: $n_{B}(j,k;t=0)=N_{B}/(L_{x}L_{y})$, with $j=1,\cdots,L_{x}$ and $%
k=1,\cdots,L_{y}$. The highlight here is the possibility of obtaining a
stationary state with two-lane ordered state if initial condition defined
above is capable to induce such similar stationary state in $B$. We solve
the problem using MC simulations and numerical integration of Eq. (\ref%
{Eq:recurrence_equations}) to compare the stages of evolution. In Fig. \ref%
{Fig:Snapshots_MC}, we show the frames for six different instants of time.
The blue (red) color means higher (lower) concentrations of a given species.

where these figures blue means more concentrated while red less concentrated

\begin{figure}[htbp]
\begin{center}
\includegraphics[width=1.0\columnwidth]{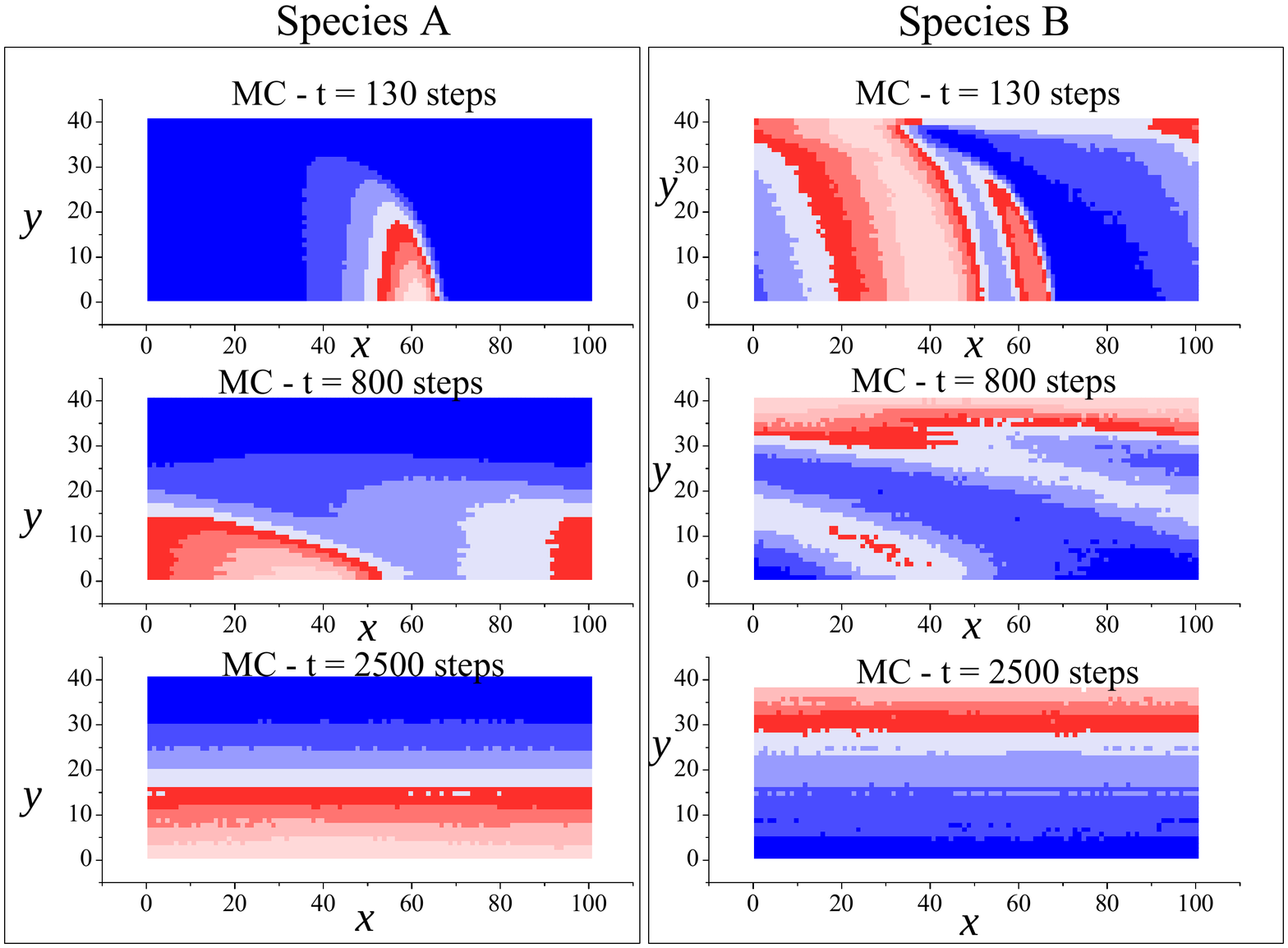}
\end{center}
\caption{Frames of some instants of time using MC simulations. We can
observe stationary states with two well distinct emergent ordered lanes .
The blue (red) color means higher concentrations of the species $A$ ($B$) in
the plots of the left (right). An assimetry is initially considered for
species $A$ according to Eq. (\protect\ref{Eq.assymetry}) while the species $%
B$ is uniformly distributed.}
\label{Fig:Snapshots_MC}
\end{figure}

\begin{figure}[htbp]
\begin{center}
\includegraphics[width=1.0\columnwidth]{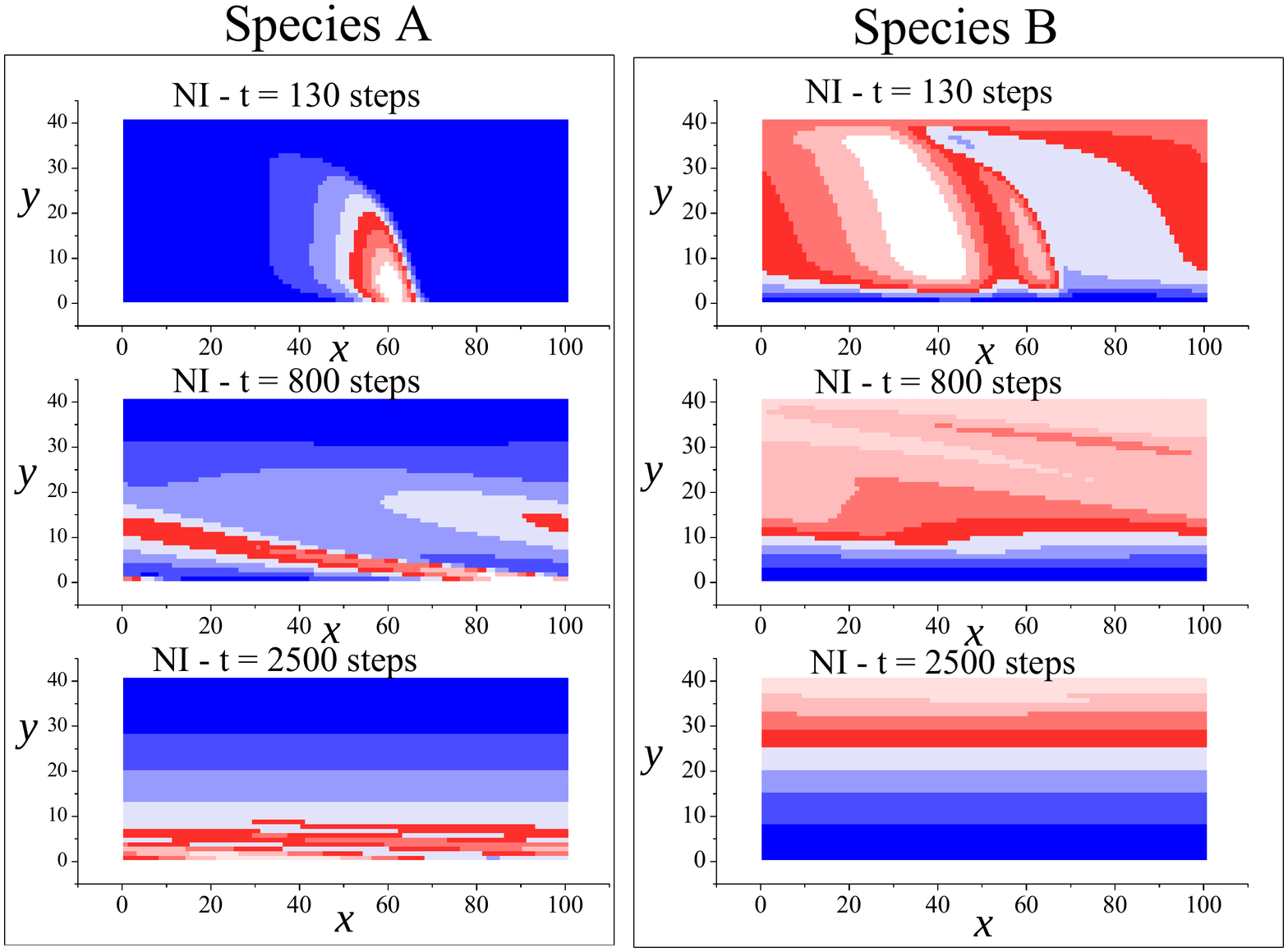}
\end{center}
\caption{Frames obtained via numerical integration of recurrence relations
for six different instants of time (as in Fig. \protect\ref{Fig:Snapshots_MC}%
). The blue (red) color means higher concentrations of the species $A$ ($B$)
in the plots of the left (right). An assimetry is initially considered for
species $A$ according to Eq. (\protect\ref{Eq.assymetry}) while the species $%
B$ is uniformly distributed.}
\label{Fig:Snapshots_EDP}
\end{figure}

We can observe that even with an asymmetric initial condition, the system
has the tendency to reach a stationary order formed by two ordered lanes
differently of what happens when one considers initial conditions where the
particles of both species are uniformly distributed. Figure \ref%
{Fig:Snapshots_EDP} shows the frames for the same instants of time when one
takes into consideration the numerical integration of Eq. (\ref%
{Eq:recurrence_equations}). Both methods lead to stationary states with two
emergent ordered lanes. It is important to observe that numerical
integration of equations corresponds to the mean field regime of MC
simulations and, although we have similar evolution and the same stationary
state, the relaxations are not synchronized. This is exactly what should
occur in Statistical Mechanics when, for example, we solve the Ising model
by performing MC simulations and via Mean-field approach. In these cases,
the critical exponents associated to possible phase transitions are
different. However, we expect that $\Phi_{\parallel}\rightarrow 1$ in both
cases even though with different speeds since Figs. \ref{Fig:Snapshots_MC}
and \ref{Fig:Snapshots_EDP} present different time scales of relaxation
dynamics.

\begin{figure}[htbp]
\begin{center}
\includegraphics[width=%
\columnwidth]{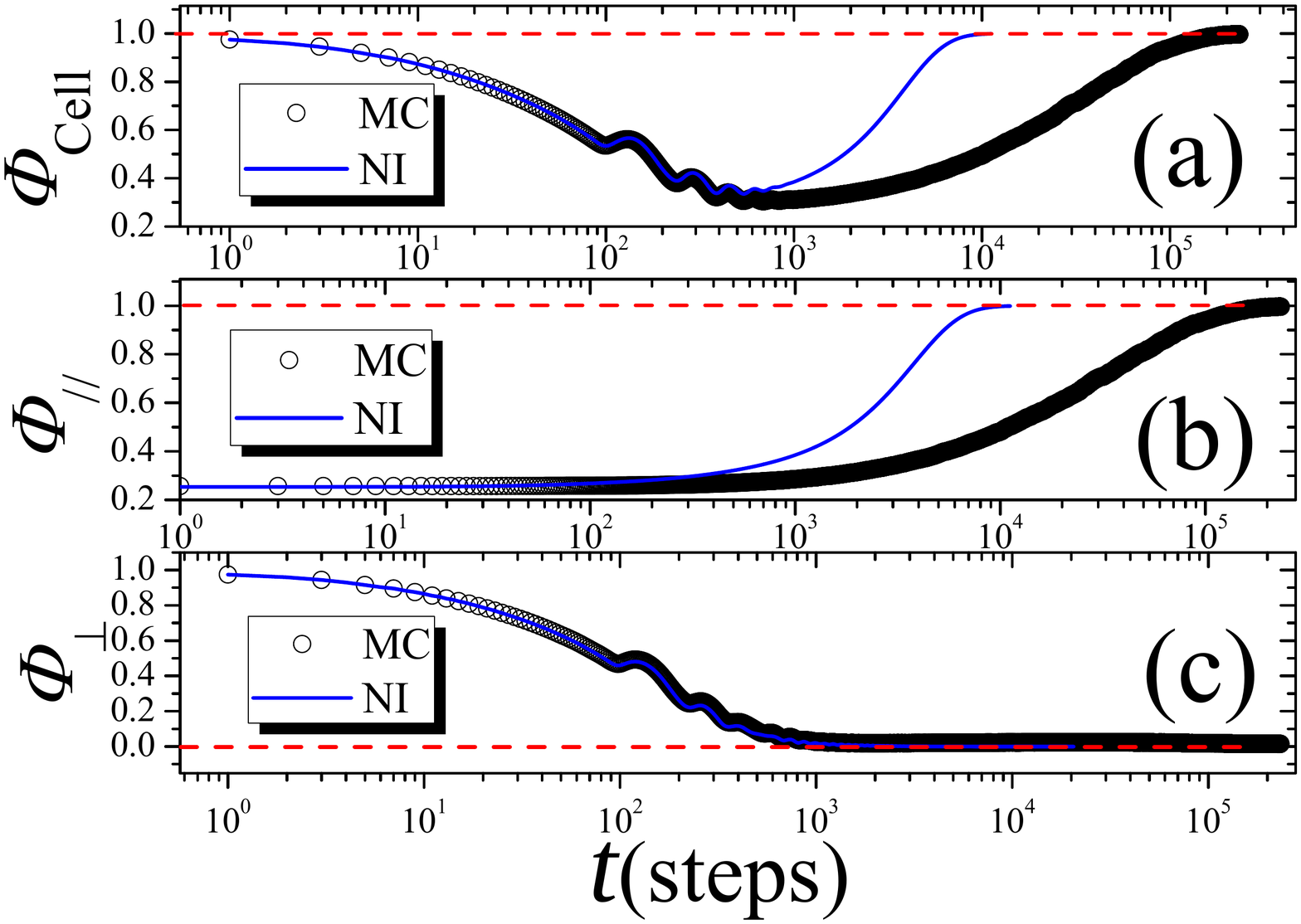}
\end{center}
\caption{Time evolution of the order parameters considering that the species 
$A$ is asymmetrically distributed in an inital vertical stripe according to
Eq. (\protect\ref{Eq.assymetry}) and the species $B$ is uniformily
distributed in the lattice. Both methods (MC simulations and NI) lead to
stationary states as previously shown in Figs. \protect\ref{Fig:Snapshots_MC}
and \protect\ref{Fig:Snapshots_EDP} composed by two emergent ordered bands
which are quantitatively corroborated by the order parameters $\Phi
_{cell}\rightarrow 1$ (a) and $\Phi _{\parallel }\rightarrow 1$ (b), when $%
t\rightarrow \infty $. We also show the behavior of $\Phi _{\perp }$ as
function of time. In this case $\Phi _{\perp }\rightarrow 0$ since the
particles do not present order by bands perpendicular to the field.}
\label{Fig:Evolution_order_parameter_assimetrico}
\end{figure}

Figure \ref{Fig:Evolution_order_parameter_assimetrico} shows the time
evolution of the three order parameters of the model. These evolutions are
obtained when carrying out the study corresponding to the Fig. \ref%
{Fig:Snapshots_MC} (MC simulations) and Fig. \ref{Fig:Snapshots_EDP} (NI).

We can observe in Fig. \ref{Fig:Evolution_order_parameter_assimetrico} that $%
\Phi_{cell}\rightarrow 1$ and $\Phi_{\parallel}\rightarrow 1$ (using MC
simulations and also NI) show the stationary case formed by two main bands
corresponding to the two species. The observed difference in the plots are
acceptable since we do not expect the same speed in the two solutions but
instead we expect the reproduction of the same steady states. However it is
important to observe that the initial part is exactly the same for all order
parameters and only the conduction to the steady state is slower for the MC
simulations. Such problem shows a interesting situation where an initial
order of species $A$ is broken after a transient regime reaches an ordered
dynamics.

\section{Summary and conclusions}

\label{Sec:Conclusions}

We have considered a two-dimensional system composed by particles of two
species, $A$ and $B$, which are able to move or not. If one of the species
remains still, it acts as fixed obstacles to the other species which is
moving. In this case, we used MC simulations and concentrated our analysis
in the statistics about crossing times (which means the needed time for one
particle cross the lattice) for different concentrations of obstacles.
However, if both species are moving, then they move in opposite diretions.
In this case, we looked into the properties of distillation times of the
particles arranged randomly in the lattice at the beginning of the
evolution, as well as the formation of ordering patterns considering
periodic boundary conditions of particles. Both studies were performed by
using MC simulations and numerical integration of recurrence relations. In
the first study, we showed that an interesting transition between the
Gaussian, power law, and exponential behavior is observed for the crossing
times of the particles, which is supported by high kurtosis values. We also
showed that the correlation times are not affected by lateral transitions.
In the second study, when both species can move, we observed that the
enlarge of lateral transitions decreases the distillation times and the
particles are more easily separated.

Finally, we observed characteristic ordering patterns, which appear also in
charged colloids motions and pedestrian dynamics, when our two-dimensional
model took into consideration periodic boundary conditions. It is important
to notice that Dickman \cite{Dickman2001} has observed that similar
spontaneous spatial ordering, which appears in systems of self-propelled
agents, can also appear in driven lattice-gas. In this case, it was found an
interesting phenomena where the drive can provoke jamming, and thereby, a
sharp reduction in the current, quite contrary to the usual relation between
bias and current. As future works, we will explore exclusion effects, for
example by introducing in our model, cells with limited maximum number of
particles and analyse these effects into ordering formation.

\textbf{Acknowledgments --} This research work was in part supported
financially by CNPq (National Council for Scientific and Technological
Development).

\end{document}